\pdfoutput=1
\documentclass[12pt]{fairmeta}

\title{An open dataset of gas adsorption in deformable metal–organic frameworks for direct air capture}

\author[1,*]{Anuroop Sriram}
\author[2,*]{Logan M. Brabson}
\author[2,*]{Xiaohan Yu}
\author[2]{Sihoon Choi}
\author[5]{Kareem Abdelmaqsoud}
\author[4]{Elias Moubarak}
\author[4]{Pim de Haan}
\author[4]{Sindy L\"owe}
\author[4]{Johann Brehmer}
\author[1]{Muhammed Shuaibi}
\author[5]{John R. Kitchin}
\author[4]{Max Welling}
\author[1]{C. Lawrence Zitnick}
\author[1,\dag]{Zachary Ulissi}
\author[2,\dag]{Andrew J. Medford}
\author[3,\dag]{David S. Sholl}

\affiliation[1]{FAIR at Meta}
\affiliation[2]{School of Chemical and Biomolecular Engineering, Georgia Institute of Technology}
\affiliation[3]{University of Tennessee-Oak Ridge Innovation Institute, Oak Ridge National Laboratory}
\affiliation[4]{CuspAI}
\affiliation[5]{Department of Chemical Engineering, Carnegie Mellon University}

\contribution[*]{Equal contribution}
\contribution[\dag]{Corresponding author}

\abstract{
Engineering economical and scalable processes for direct air capture (DAC) from humid air relies on atomistic modeling to search and identify promising sorbent materials from a vast chemical space. Machine learning has emerged as a powerful tool for predicting material properties in an efficient yet accurate manner, but these models require large sets of carefully curated training data. We present the Open DAC 2025 (ODAC25) dataset comprising nearly 60 million DFT single-point calculations for \ce{CO2}, \ce{H2O}, \ce{N2}, and \ce{O2} adsorption in $\sim$15,000 MOFs. ODAC25 introduces functionalized MOFs, high-energy adsorabte placements, and synthetically generated frameworks to maximize the versatility of models trained on the dataset. ODAC25 improves upon the accuracy of DFT calculations in previous works and applies engineering intuition from recent work regarding the treatment of adsorbate-induced MOF flexibility. We release state-of-the-art machine-learned interatomic potentials trained on ODAC25 and demonstrate their utility via Henry's coefficient predictions. ODAC25 enables MOF screening for DAC at a resolution previously unattainable.
}

\correspondence{A.S. \email{anuroops@meta.com}, A.J.M (ajm@gatech.edu), D.S.S (shollds@ornl.gov)}

\metadata[Code]{\url{https://github.com/facebookresearch/fairchem}}
\metadata[Data]{\url{https://huggingface.co/facebook/ODAC25}}

\newcommand{\carbondioxide}[0]{\ce{CO2}}
\newcommand{\water}[0]{\ce{H2O}}
\newcommand{\nitrogen}[0]{\ce{N2}}
\newcommand{\oxygen}[0]{\ce{O2}}

\newcommand{\ev}{\text{eV}}
\newcommand{\angstrom}{\text{\normalfont\AA}}

\newcommand{\eva}{\ev/\angstrom}

\newcommand{\emae}{EMAE$\downarrow$}
\newcommand{\fmae}{FMAE$\downarrow$}

\newcommand{\runinhead}[1]{\textit{#1}\ }
\usepackage{titlesec}
\titleformat{\subsection}
  {\normalfont\itshape\large}{\thesubsection}{1em}{}
\usepackage{caption}
\DeclareCaptionLabelSeparator{myperiod}{.}
\DeclareCaptionFormat{myformat}{#1#2 #3}
\usepackage[version=4]{mhchem} %
\usepackage{booktabs}
\usepackage{siunitx} %
\usepackage{tablefootnote}
\usepackage{adjustbox}
\usepackage{listings}
\usepackage{changepage} %

\usepackage{threeparttable}
\usepackage{multirow}
\usepackage{mathtools}

\usepackage{tabularx}
\usepackage[percent]{overpic}

\DeclarePairedDelimiter\ceil{\lceil}{\rceil}

\lstdefinelanguage{ini}{
    basicstyle=\scriptsize,
    morecomment=[l]{;},
    morecomment=[l]{\#},
    morestring=[b]",
    moredelim=[s][\color{blue}]{[}{]},
    alsoletter={=},
}

\begin{document}

\maketitle

\setcounter{section}{0}
\textit{Notice of Copyright: This manuscript has been authored by UT-Battelle, LLC, under contract DE-AC05-00OR22725 with the US Department of Energy (DOE). The publisher acknowledges the US government license to provide public access under the DOE Public Access Plan (\url{http://energy.gov/downloads/doe-public-access-plan}).}

\section{Introduction}
\label{sec:intro}

Direct air capture (DAC) is a promising carbon capture technology for addressing global climate change through negative emissions~\cite{Sood2017} and is among the most important chemical separations for a sustainable future \cite{Sholl2016}. DAC is an active area of research spanning atomistic simulations for sorbent discovery \cite{Findley2021,Lim2025,Daglar2020,Colon2014} to process optimization and life cycle analyses \cite{Prisma, Wiegner2022,Kim2022}. Unlike traditional point-source capture, DAC facilities can operate at ambient conditions with fewer geographical constraints~\cite{SanzPrez2016}.
However, most existing DAC sorbents require energy-intensive regeneration that increases costs and reduces environmental benefits~\cite{Tiainen2022}.
Metal–organic frameworks (MOFs)~\cite{Furukawa2013} offer a promising alternative as highly tunable, modular porous materials with potential for low-temperature sorbent regeneration~\cite{Peedikakkal2020, Jamdade2023}. Given the vast chemical space and synthesis challenges of MOFs~\cite{Moosavi2019, Moosavi2020, Stock2012}, high-throughput computational screening (HTCS) has become essential for developing better sorbents~\cite{Daglar2020, Wilmer2012}. Identifying promising sorbents in this manner requires models trained on large amounts of accurate and carefully curated data.

The Open DAC 2023 (ODAC23) dataset~\cite{ODAC23} introduced over 38 million DFT calculations for \carbondioxide~and \water~adsorption across 8,400 MOFs, identifying interesting candidates for DAC and the influence of key chemical motifs such as open-metal sites, parallel aromatic rings, metal-oxygen-metal bridges, and uncoordinated nitrogen atoms.
Prior HTCS studies relying on classical force fields~\cite{Harrison2018} such as UFF(4MOF)~\cite{UFF, UFF4MOF1, UFF4MOF2} and rigid framework assumptions often failed to identify viable materials~\cite{Findley2021, Colon2014, Lee2021, Qiao2016}.
By incorporating and DFT-level accuracy, ODAC23 identified MOF sites with potential \carbondioxide~selectivity that classical approaches missed.
While DFT calculations are computationally expensive for large-scale screening, machine learning force fields (MLFFs) trained on this DFT data demonstrated the promise of approaching this level of accuracy while dramatically accelerating high-throughput screening.

Despite its advances, ODAC23 had limitations. First, it was limited to two adsorbates, \carbondioxide~and \water, while realistic air separations require modeling \nitrogen~and \oxygen~as well. Second, ODAC23 did not explore functionalization of MOF linkers or open metal sites (OMSs) \cite{Cohen2012, Cai2020}, approaches that offer significant potential to enhance \carbondioxide~selectivity while reducing regeneration energy~\cite{Chong2017, Gu2022}. Third, ODAC23 reported only adsorption energies relative to relaxed empty MOF structures, potentially introducing artifacts when guest molecules stabilized MOFs into lower-energy configurations than the empty framework reference state. These limitations, combined with ongoing challenges around MOF structural integrity in computational studies~\cite{White_MOSEAC, jin2025correspondence}, motivated the development of a more comprehensive and methodologically rigorous dataset.

Recent advances in MLFFs have enabled accurate prediction of molecular and materials properties at reduced computational cost. However, large-scale screening for DAC presents challenges including low \ce{CO2} concentrations, competing gases (\ce{N2}, \ce{O2}, and \ce{H2O}), and heterogeneous binding environments, requiring MLFFs to generalize across diverse framework topologies, adsorbates, and configurations.

To address these limitations, we introduce in this paper the Open Direct Air Capture 2025 (ODAC25) dataset comprising nearly 60 million DFT calculations across 15,000 MOFs with four adsorbates: \ce{CO2}, \ce{N2}, \ce{O2}, and \ce{H2O}.
ODAC25 substantially expands ODAC23 in terms of scale, diversity, and computational accuracy. We systematically improved the accuracy of all calculations by performing various MOF validation checks, correcting for systematic errors introduced by incompletely converged k-point sampling, and re-relaxing each bare MOF structure to account for adsorbate-induced MOF deformation. ODAC25 enables more robust MLFF development by including functionalized MOFs with both linker and open-metal site functionalization, synthetically generated frameworks that extend structural and chemical diversity beyond what is available in experimental databases, as well as high-energy multi-component adsorption configurations derived from Grand Canonical Monte Carlo (GCMC) simulations. We also highlight the important role of adsorbate-induced deformation in building MOF datasets. Deformation significantly affected our results and represents a major scientific advancement over existing literature ODAC25 is thus designed not only to improve MLFF performance, but also to support realistic benchmarking and thorough screening of a wider range of MOF chemistries for DAC.

We also release state-of-the-art MLFFs (EquiformerV2~\cite{liao2023equiformerv2}, eSEN~\cite{fu2025learningsmoothexpressiveinteratomic}, and UMA~\cite{Wood2025Uma}) trained on ODAC25 and benchmarked on energy, force, and MOF deformation predictions. We evaluate the MLFFs for predicting Henry's law coefficients via Widom insertions---a challenging step representative of downstream adsorption isotherm simulations. The ODAC25 dataset and models provide the most complete tool set to date for sorbent identification, atomistic modeling, and eventual DAC process modeling and life cycle analysis.

\setcounter{section}{1}
\section{Results}
ODAC25 introduces several improvements that can be broadly categorized into two groups: improvements to DFT calculation accuracy (\cref{subsec:accuracy_improvements}) and improvements to the diversity of the dataset (\cref{subsec:diversity_improvements}).

\label{sec:dataset_structure}

\subsection*{Accuracy and Data Quality}
\label{subsec:accuracy_improvements}
\runinhead{Validation of MOF structures.}\label{subsubsec:validation}The first improvement introduced by ODAC25 addresses the chemical validity of the dataset's MOFs. White et al.~\cite{White_MOSEAC} suggested that high rates of structural errors exist in some MOF databases by applying an algorithm based on semi-empirical calculations to check metal oxidation states. Jin et al.~\cite{D5DD00109A} released an algorithm to validate and correct MOF structural files called MOFChecker. To mitigate concerns related to MOF structural accuracy, we performed several checks on all structures in ODAC25 using MOFChecker v0.9.6~\citep{D5DD00109A}. \Cref{tab:mofcheck} shows the checks performed and the percentage of ODAC25 structures that fail each. Jin et al. also screened ODAC MOFs for net charges according to stoichiometry and metal oxidation state predictions built into MOFChecker v2, and those results are available in their report \cite{jin2025correspondence}.

We note that the validity of semi-empirical oxidation states and similar measures to assess the quality of MOFs is unclear. The MOF structures in ODAC25 are fully relaxed with DFT calculations in charge-neutral periodic cells. Atomic point charges can be assigned from the electron distribution in these DFT calculations using DDEC charges \cite{ManzDDEC} and related methods \cite{EQeq_method, Ongari2019}. For these reasons, we retained MOFs flagged as problematic by MOFChecker, and users may choose either the ``\textit{filtered}'' or ``\textit{full}'' dataset depending on their application. All discussion in this work is for the ``\textit{full}'' (unfiltered) dataset.

\runinhead{Improving convergence in reciprocal space for DFT calculations.}\label{sec:kpoint_correction}DFT calculations in our previous work used a \( 1 \times 1 \times 1 \) k-point sampling for all MOFs, which can potentially cause numerical convergence issues for MOFs with small unit cells. A more accurate approach is to set the number of k-points to $\ceil*{K/a} \times \ceil*{K/b} \times \ceil*{K/c}$ for a unit cell of size $a \times b \times c$ for a suitably large \textit{k}-point density, $K$. \Cref{fig:convergence_errors_correction}a shows the k-point convergence for 100 randomly selected systems from ODAC23. Around 7\% of the calculations with the ODAC23 settings have noticeable systematic errors ($>$ 0.2 eV) as compared to $K=40$ \AA{}.

Re-running full DFT relaxations at a higher k-point density is computationally expensive, as each relaxation trajectory contains hundreds of frames. We instead use a simple correction method: numerical tests confirmed that energy errors due to low k-point density remain nearly constant across all frames within a trajectory (\cref{fig:convergence_errors_correction}b). We can therefore upgrade calculations by computing the energy errors of the initial and final frames and using their average to correct all frames in the trajectory. This reduces convergence errors in total energy by an order of magnitude, to $\sim 0.01$ eV.
Since the average trajectory consists of over 200 frames, this procedure incurs less than 1\% of the computational cost of na\"{\i}vely re-running each frame. We did not apply these corrections to forces because we found the force errors to be very small ($\sim 0.01~\eva$). \Cref{tab:dft_settings} provides the full DFT settings used in this work.

\runinhead{Re-relaxations of empty MOFs.}\label{sec:rerelaxations}Nearly all previous HTCS studies of MOFs approximated MOF structures as being rigid. Using rigid MOFs means that adsorption energies are equivalent to the sum of host-guest interactions. By performing DFT relaxation for a large number of MOFs, ODAC23 provided interesting insight into the influence of adsorbed molecules on MOF structures. Although the presence of adsorbates causes (typically local) deformation in MOFs, it might be expected that removing the adsorbate and re-relaxing the MOF would lead to the same structure as the original empty MOF. In the ODAC23 dataset, however, we found many instances where this kind of re-relaxation yields a more energetically favorable empty MOF than the original structure due to perturbations and broken symmetries induced by the adsorbate. The energy of the empty MOF is an important quantity in computing molecular adsorption energies, so failure to use the correct ground state empty MOF can cause significant artifacts in the adsorption energies \cite{Brabson2025,D5DD00109A}.

To address this effect, we adopted the treatment of adsorbate-induced deformation outlined in recent work \cite{Brabson2025}. We re-relaxed the atomic positions of empty MOFs after every converged MOF+adsorbate DFT relaxation in ODAC25. Only local distortions caused by slight changes over all the unit cell's atomic positions were considered here. Unit cell relaxation was not considered, so breathing behaviors like that observed in MIL-53 were not captured \cite{Weser2017}. Because we typically sampled multiple adsorbate identities and placements in each MOF, these re-relaxations potentially generated more than one structure for each MOF. In all calculations below that require a reference energy for an empty MOF, the lowest energy structure among the available collection of re-relaxed MOFs is used. Thus, our adsorption energies in ODAC25 combine host-guest interactions with adsorbate-induced deformation based on the principle that the physically correct empty MOF configuration is the most energetically favorable one regardless of which guest molecule was used to obtain said configuration.

\begin{figure*}[htbp]
    \centering

    \begin{subfigure}[b]{0.49\textwidth}
        \centering
        \begin{overpic}[width=\textwidth]{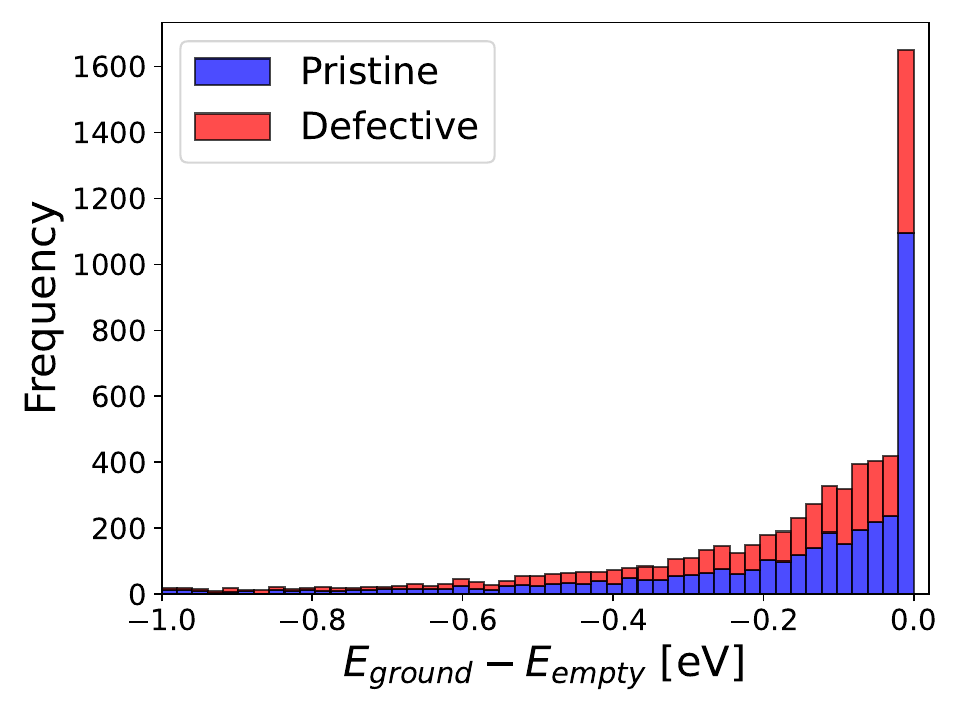}
            \put(2,85){\fontsize{20}{24}\selectfont \textbf{(a)}}  %
        \end{overpic}
        \label{subfig:deform_hist}
    \end{subfigure}
    \hfill
    \begin{subfigure}[b]{0.49\textwidth}
        \centering
        \begin{overpic}[width=\textwidth]{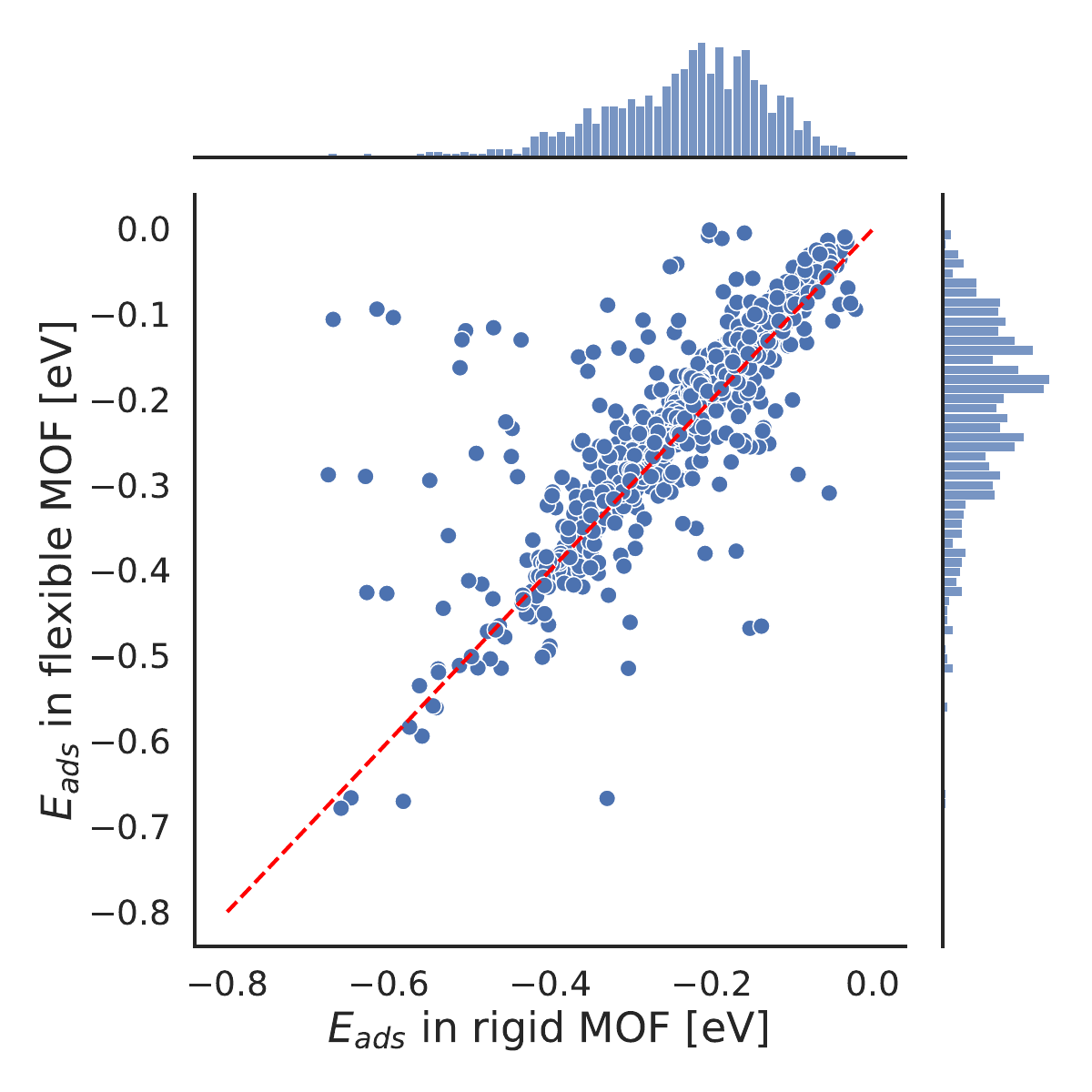}
            \put(2,85){\fontsize{20}{24}\selectfont \textbf{(b)}}  %
        \end{overpic}
        \label{subfig:rigid_vs_flex}
    \end{subfigure}

    \caption{Magnitude and effects of MOF deformation on DFT adsorption energies. (a) Comparison of original DFT-relaxed empty MOF energies ($E_{empty}$) to the most favorable MOF energy found across all ODAC25 relaxations ($E_{ground}$) for 3,592 pristine and 2,788 defective ODAC23 MOFs. (b) Comparison of DFT adsorption energies in rigid and flexible MOFs for 976 MOF+adsorbate configurations.}
    \label{fig:rerelax}
\end{figure*}

\Cref{fig:rerelax}a shows energy differences between the most favorable empty MOF geometry in ODAC25 ($E_{ground}$) and the original DFT-relaxed empty MOF in ODAC23 ($E_{empty}$), both corresponding to DFT local minima. $E_{ground}$ is determined from the minimum energy MOF configuration resulting from adsorption of \ce{N2} and \ce{O2}, as discussed in \cref{sec:new_adsorbates}, or from any combination of \ce{CO2} and \ce{H2O} used in ODAC23. Many MOFs show non-trivial energy differences. Positive adsorption energies can result even when host-guest interactions are favorable, because the empty MOF reference energy is more energetically favorable than that which would be used under a rigid MOF assumption.

In ODAC25, we corrected all ODAC23 adsorption energies and performed all additional calculations (\cref{subsec:diversity_improvements}) using this method. Our ODAC25 adsorption energy calculations featuring this method supersede those presented in ODAC23. We note that for calculations that rely on total energies such as the training of MLFFs, all of the distinct local minima obtained by DFT for a given structure are useful. This re-relaxation approach is only necessary to calculate physically relevant adsorption energies.

\Cref{fig:odac23_rerelax_dist} shows the distributions of ODAC23 adsorption energies and their corresponding ODAC25 adsorption energies corrected using MOF re-relaxations. The adsorption energies after accounting for re-relaxation (orange data in \cref{fig:odac23_rerelax_dist}) are shifted towards less favorable adsorption than the adsorption energies reported with ODAC23 (blue data). The adsorption energy median shift is $>0.1$ eV for all splits. Nevertheless, there are many examples in which the adsorption energy for individual \ce{CO2} or \ce{H2O} molecules is chemisorption-like (that is, more favorable than –0.5 eV). Using the minimum MOF energy in this manner is the physically relevant quantity, and ODAC25 adsorption energies should be used instead of the ODAC23 energies.

Our treatment of adsorbate-induced deformation is a significant departure from the rigidity assumption used in most studies and results in a systematic shift towards weaker binding energies. To quantify the importance of deformation, we computed DFT adsorption energies in 976 MOF+adsorbate configurations using the rigidity assumption. We selected configurations where the flexible MOF approaches the global minimum to within 0.05 eV, avoiding strongly positive adsorption energies caused by discovering new favorable configurations. \Cref{fig:rerelax}b compares the DFT adsorption energies in rigid MOFs to those in deformable MOFs as presented in our dataset. In 15.8\% of configurations, the rigid and flexible adsorption energies differ by more than 0.05 eV, and this difference grows to more than 0.1 eV in 5.9\% of configurations. \Cref{fig:COJHIT_rigid_vs_flex} shows one such example featuring \ce{H2O} adsorption in COJHIT where the DFT adsorption energies in the rigid and flexible MOF differ by 0.337 eV. The rotation of the linker upon adsorption is also visibly clear. Deformation affects the DFT adsorption energy in a significant number of cases and thus should not ignored in MOF modeling work.

Whether MOFs identified as promising under a rigidity assumption remain promising when modeled as deformable is an open question. Recent work has begun integrating sorbent discovery across scales, from atomistic modeling to process optimization, but these efforts rely on many assumptions and remain a developing area of research \cite{Prisma,Cleeton2023}. ODAC25 focuses on the high-fidelity atomistic modeling that serves as the foundation for all such downstream efforts, since the unreliability of classical force fields as DFT surrogates is well-documented \cite{Findley2021,Brabson2025}.

\subsection*{Diversity of Adsorbates, Adsorbents, and Energetics}
\label{subsec:diversity_improvements}
\runinhead{New adsorbates.}\label{sec:new_adsorbates}ODAC25 includes two new adsorbates, \ce{N2} and \ce{O2}, to enhance the dataset’s coverage of situations relevant to modeling DAC. Adsorbates are placed within the DFT-relaxed MOF unit cells using Monte Carlo sampling with classical force fields to identify diverse, energetically favorable configurations, then relaxed using DFT in calculations that allow all atoms to move with the PBE+D3 functional. ODAC25 contains nearly 56K \ce{N2} and \ce{O2} relaxation trajectories across $\sim6,400$ pristine and defective ODAC23 MOFs. \Cref{fig:odac23_n2o2_dist} summarizes the \ce{N2} and \ce{O2} adsorption energy results.

\begin{figure*}[ht!]
    \centering
    \includegraphics[width=0.95\textwidth]{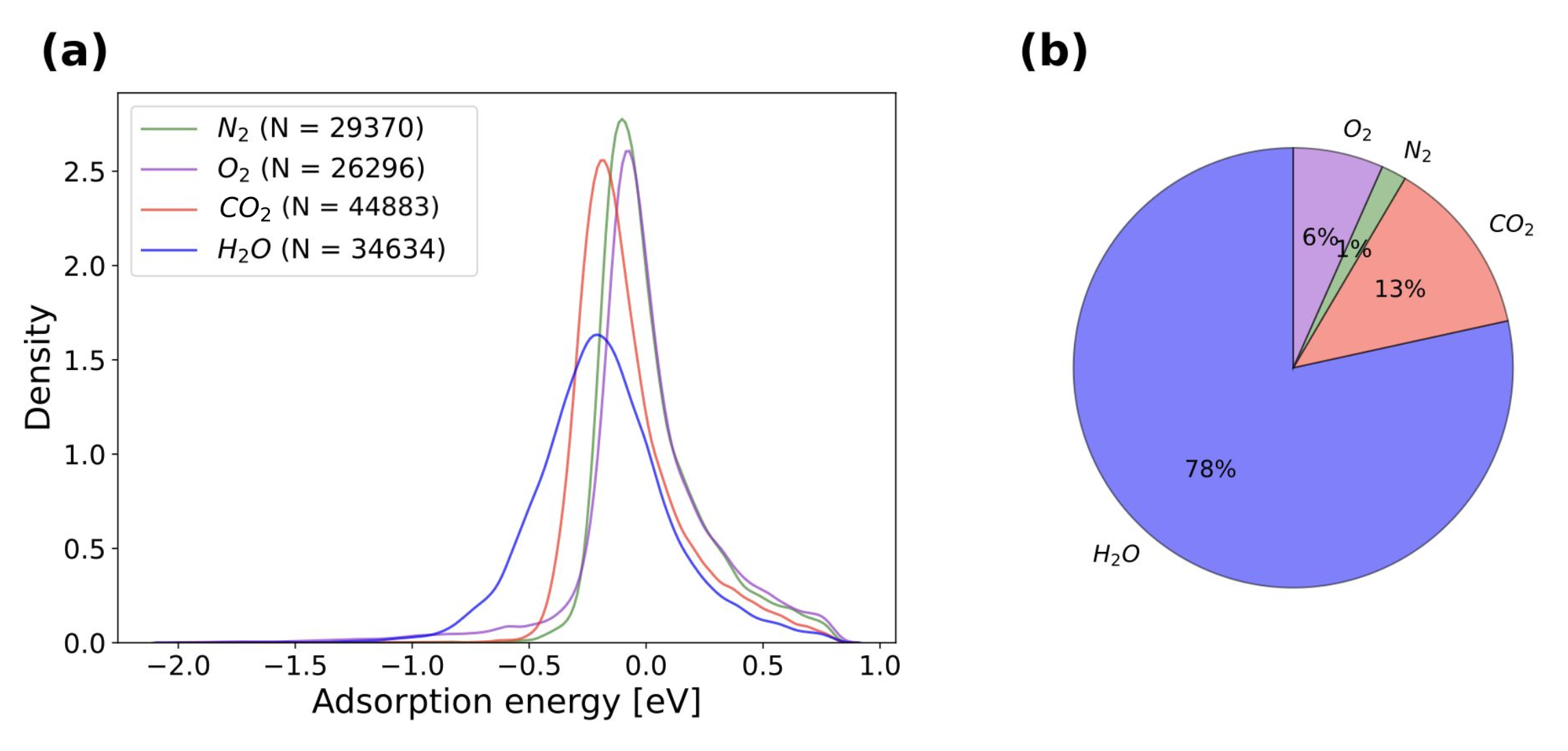}
    \caption{Summary of \ce{N2} and \ce{O2} adsorption statistics. (a) Distribution of adsorption energies for different adsorbates computed using kernel density estimation. (b) Percentage of MOFs that adsorb each adsorbate most strongly, determined by taking the strongest adsorption energy of each adsorbate across all sampled active sites in a given MOF framework.}
    \label{fig:combined_adsorption}
\end{figure*}

The inclusion of multiple adsorbates with our DFT calculations can provide information about effects including competitive adsorption, redox activity, and the impact of \ce{O2} reactivity on \ce{CO2} adsorption sites. \Cref{fig:combined_adsorption}a shows the distribution of binding energies from each DFT-relaxed configuration with a single adsorbed molecule (referenced to the most energetically favorable empty MOF structure available, as described above). \Cref{fig:n2o2_hist} shows the histogram corresponding to the kernel density plot in \cref{fig:combined_adsorption}a. Across the entire ODAC25 dataset for single-adsorbate configurations, 30.8\% of the configurations have positive adsorption energies, meaning adsorption is not energetically favored, 63.5\% of the structure have adsorption energies between -0.5 and 0 eV, corresponding to physisorption, and 5.7\% (8,320 configurations) show stronger adsorption in the chemisorption regime. Capturing chemisorption, which may enable cooperative adsorption beneficial for DAC \cite{Zhu2024}, is only possible using DFT instead of classical FFs.

We find 167 ODAC25 MOFs where the strongest observed adsorption energy is with \ce{N2}. Our calculations sampled only a small number of configurations for each molecule (typically 2-5), so some of these cases may stem from incomplete sampling of the full potential energy surface for adsorbates. A full exploration of selectivity requires the Widom insertion method as discussed in \cref{sec:henrys_coeff}. A small number of example of metal-organic clusters exhibiting \ce{N2} chemisorption have been reported previously \cite{Zhao2021,Zhang2023,Demir2019,Jaramillo2020}, and we find similar cases featuring transition metal sites in ODAC25. Figure \ref{fig:n2_chemisorption} shows one such example in the MOF with CSD code DEJRUH. Bader charge analysis shows charge transfer indicative of chemisorption, with the cobalt site donating 0.25 e and the nearest nitrogen atom accepting 0.31 e. Our use of DFT in ODAC25 enabled us to find such interesting chemistries that would  be missed by classical force fields.

Accurately describing \ce{N2} and \ce{O2} adsorption near redox-active metal sites is challenging using non-hybrid density functional approximations (DFAs). Prior study of DFAs for describing redox-dependent binding at MOF OMSs showed that PBE artificially disfavors high-spin states due to strong many-electron self-interaction errors \cite{Rosen2020}. The resulting overprediction of binding energy strength can be mitigated by the inclusion of a Hubbard \textit{U} correction \cite{HubbardU}. Although Hubbard corrections are computationally cheap, selection of appropriate \textit{U} parameters across a large and diverse dataset is non-trivial \cite{Yu2020_Hubbard}.

Spin presents additional challenges for describing \ce{N2} and \ce{O2} adsorption, especially for \ce{O2} chemisorption near open metal sites. Spin effects have been shown to be instrumental in \ce{O2} binding on transition metal complexes \cite{Sun2020_spin} and in MOFs \cite{Jose2022}. DFT-computed binding energies for small molecules can be highly sensitive to the initial spin configuration when adsorption occurs near redox-active metal sites \cite{You2018,Jaffe2020,Rosen2020_redox}. Strong \ce{O2} adsorption energies may also result from \ce{O2} dissociation. Screening our dataset for O-O bonds greater than 1.4 Å revealed 619 cases out of 38,441 where \ce{O2} dissociation occurred. This decreases to 226 when defining \ce{O2} dissociation as an O-O bond length of greater than 1.5 Å.

ODAC25 prioritizes consistent DFT settings across all calculations and does not attempt to address shortcomings of the PBE functional or the effects of spin polarization. Enumerating all plausible spin states at this scale is intractable, although recent advances in spin-informed MLFFs may provide a route forward \cite{10.1073/pnas.2422973122}. All DFT calculations in ODAC25 were spin-polarized with initial magnetic moments set to the default 1.0 $\mu_B$ for all atoms. Given these limitations of the underlying DFT, results from ODAC25 models, particularly for \ce{O2}, should be used judiciously. Additional ab initio calculations should be used to more fully explore adsorption behavior near redox-active metal sites in MOFs of particular interest. We emphasize, however, that our DFT still far exceeds the accuracy provided by classical FFs despite its shortcomings, especially for chemisorption cases.

\begin{figure}[htbp]
\centering
\includegraphics[width=0.95\textwidth]{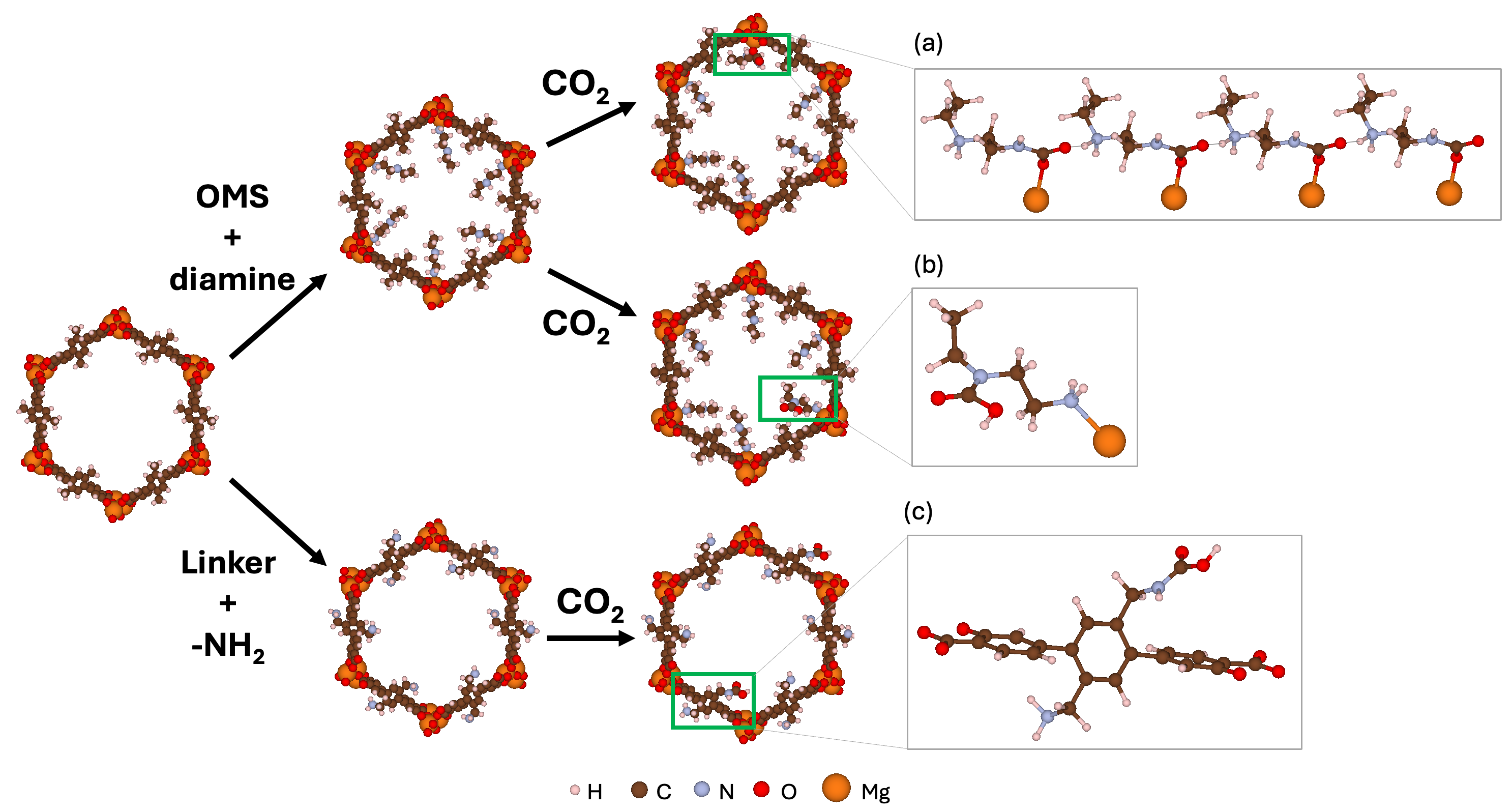}
\caption{Two approaches (OMS and linker functionalization) used to generate functionalized MOFs using N-Ethylethylenediamine (een) in IRMOF-74-III (CSD code RAVWAO) as an example. (a) OMS functionalization with a metal-oxygen bond. (b) OMS functionalization with a metal-nitrogen bond. (c) Linker functionalization.}\label{fig_mechanism}
\end{figure}

\runinhead{New adsorbents.}To broaden the scope of MOFs included in our dataset, we generated new MOF structures using two MOF amine functionalization methods. We first used linker functionalization, which has been shown experimentally to enhance \ce{CO2} adsorption~\cite{linker_Popp_yaghi_2017,linker_Fracaroli_Yaghi_2014,linker_Ethiraj_Bordiga_2014,linker_Chen_Cheng_2015,linker_Singh_Nagaraja_2017,linker_Rada_Abid_Sun_Wang_2015}. We generated amine-functionalized MOFs using seven organic linkers with amine groups that were previously used in MOF synthesis (see \cref{tab_linkers}). Second, we generated structures using open metal site (OMS) functionalization with diamines in which one amine is bound to an open metal site, while the other remains exposed in the pore. Experimental studies on functionalized MOFs of the latter kind including MIL-101~\cite{diamine_Hu_Jiang_2014,diamine_Vo_Kim_Kim_2020}, Mg-dobdc~\cite{diamine_Choi_Jones_2012,diamine_Liao_Chen_2016,diamine_Bernini_Narda_2015}, and $\mathrm{M_2(dobpdc)}$ (M=Mg, Mn, Fe, Co, Zn)~\cite{Siegelman2017, McDonald_Long_2015} have demonstrated exceptional \ce{CO2} capture ability, especially at low \ce{CO2} partial pressures. We functionalized MOF structures with different concentrations of ten diamines (\cref{tab_diamines}), including primary, secondary, and tertiary amines. An example of the functionalization process is shown in \Cref{fig_mechanism}, and full details are provided in the Methods section.

\noindent

It is interesting to compare \ce{CO2} and \ce{H2O} adsorption energies in MOFs with and without functionalization. We considered MOFs with data for adsorption of a single \ce{CO2} and a single \ce{H2O} in both the non-functionalized base MOF and at least one functionalized MOF, excluding cases corresponding to \ce{H2O} adsorption in functionalized MOFs generated by the reactive methods defined above. This gave data from 2,093 functionalized MOFs derived from 73 CoRE MOFs, with a total of 12,065 and 10,049 \ce{CO2} and \ce{H2O} adsorption energies in the functionalized MOFs, respectively. Notably, the reference energies for empty functionalized MOFs most often result from MOF + \ce{H2O} configurations since the MOFs do not deform enough in the presence of \ce{CO2} to approach the most energetically favorable empty MOF configurations (see \cref{fig:odac25_func_dist}). This causes many of the resulting \ce{CO2} adsorption energies to be positive.

\begin{figure*}[ht!]
    \centering
    \includegraphics[width=\textwidth, height=0.63\textheight, keepaspectratio]{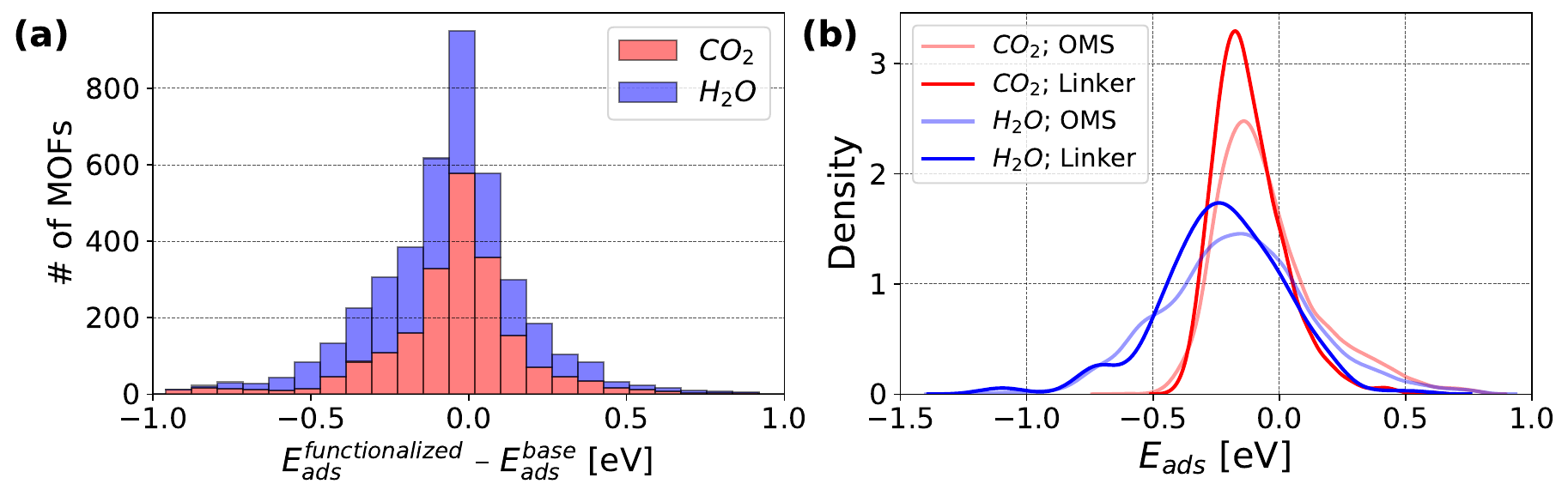}
    \caption{Effect of functionaliation on DFT adsorption energies. (a) Difference in most favorable adsorption energies between functionalized MOFs and their corresponding base MOFs. (b) Kernel density estimation plots for all \ce{CO2} and \ce{H2O} adsorption energies in linker-functionalized (N = 147) and OMS-functionalized (N = 2,672) MOFs. There are 672 \ce{CO2} and 577 \ce{H2O} adsorption energies and 11,395 \ce{CO2} and 9,472 \ce{H2O} adsorption energies in  linker- and open metal site-functionalized MOFs, respectively.}
    \label{fig:func_combined}
\end{figure*}

\Cref{fig:func_combined}a compares the most favorable \ce{CO2} and \ce{H2O} adsorption energies from the non-functionalized MOFs to the functionalized MOFs described above. Functionalization in general makes adsorption more favorable, as expected, with roughly equal effects on \ce{CO2} and \ce{H2O} adsorption energies. \Cref{fig:func_combined}b compares the adsorption energy distributions in linker-functionalized MOFs relative to OMS-functionalized MOFs. The \ce{CO2} adsorption energies are broadly similar in the two types of functionalized MOFs, which differs from existing literature showing weaker adsorption energies in linker-functionalized MOFs (–0.3 to –0.4 eV) relative to OMS-functionalized MOFs (–0.6 to –0.9 eV) \cite{Darunte_Walton_Sholl_Jones_2016}. \Cref{fig:func_type_histogram} shows the histogram corresponding to the kernel density plot in \cref{fig:func_combined}b. For OMS-functionalized MOFs, \cref{fig:diamine_kde} shows the adsorption energy distributions in MOFs functionalized with each diamine, and \cref{fig:diamine_scatter} shows the \ce{CO2} adsorption energies in OMS-functionalized MOFs as a function of diamine concentration. There is minimal correlation between diamine type, diamine concentration, and adsorption energies in our data because our adsorbate placement method did not guarantee placement near functionalized linkers or OMSs. Further analysis, including careful treatment of adsorbate placements, is needed to draw detailed conclusions about the effects of functionalization, and is outside the scope of this work.

\runinhead{High-energy configurations from GCMC.}\label{sec:gcmc_data}To further enhance the diversity of training data in ODAC25, we included DFT energies for adsorption configurations not at local energy minima generated via Grand Canonical Monte Carlo (GCMC) simulations. These configurations were generated using the RASPA package with classical force fields, employing the Universal Force Field (UFF) for the MOF frameworks and holding the MOF rigid in its initial DFT-relaxed structure from ODAC23.

From these GCMC simulations, random intermediate configurations were saved and single-point DFT calculations were performed (without energy minimization) to compute energies and forces. With this approach, we generated over 2.7 million single-point DFT calculations. \Cref{fig:gcmc_hist} shows a histogram of the number of \ce{CO2} and \ce{H2O} molecules in this data split. This large collection of DFT expands the data set in two important ways relevant for accurately predicting adsorption isotherms: inclusion of states with large numbers of adsorbed molecules and states that may differ considerably from adsorption in energy-minimized adsorption sites.

Multi-adsorbate configurations capture gas-gas interactions, though these may be more accurately treated with classical force fields given the significant degrees of freedom involved. We do not predict adsorption isotherms in this work because isotherms involving water require advanced MC sampling techniques to avoid simulations becoming stuck due to hydrogen-bonded water clustering \cite{Siderius2024,Mazur2024}, and multi-component isotherms require lengthy simulation times with MLIPs \cite{Liu2024}. Recent software such as gRASPA has made progress toward faster GCMC simulations, but interfacing with third-party MLIPs remains limited \cite{gRASPA}.

\runinhead{Synthetically generated MOFs.}\label{sec:synthetic_mofs}We further increased the diversity of ODAC25 by including 460 synthetically generated MOFs. To this end, we generated 53,000 candidate structures with CuspAI's in-house generative model of MOFs. Full details are provided in the Methods section. \Cref{fig:synthetic_mofs} shows that the resulting hypothetical structures successfully extend the space covered by ODAC25 toward more complex geometries featuring larger pores, higher surface areas, and lower densities compared to experimental structures. \Cref{fig:synthetic_dist} shows the distributions of \ce{CO2}, \ce{H2O}, \ce{N2}, and \ce{O2} adsorption energies in these synthetic MOFs. Configurations with multiple (up to 15) adsorbate molecules are excluded from \cref{fig:synthetic_dist}.

\setcounter{section}{2}
\subsection*{ML Interatomic Potentials}
\label{sec:Results}

The development of machine learned interatomic potentials (MLIP) has seen rapid progress over the last few years, and various architectures have been developed for molecules and materials~\cite{schutt2017schnet,sriram2022towards,gasteiger2021gemnet,gasteiger2022graph,schutt2021equivariant,passaro2023reducing,liao2023equiformerv2,Batatia2022mace}. More recently, a number of foundational MLIPs trained on multiple datasets across different classes of materials and molecules have been developed, demonstrating that a single model can accurately predict energies and forces across different chemical modalities~\cite{Batatia2022mace,Wood2025Uma}.
The ODAC23 paper demonstrated that state-of-the-art MLIPs trained on adsorption energies and forces from that dataset significantly outperformed classical force fields based on UFF, particularly in the chemisorption regime.

\runinhead{Adsorption energy and force evaluations.}\label{sec:energy_force_evaluation}In this section, we describe the results of recent MLIPs trained on the ODAC25 dataset on the \textit{Structure to Energy and Forces (S2EF)} task, which involves predicting the non-relaxed adsorption energy and atomic forces. This is analogous to evaluating a force field.
Our experiments use two recent model architectures: eSEN~\cite{fu2025learningsmoothexpressiveinteratomic}, and UMA~\cite{Wood2025Uma}.
Our training setup is similar to the ODAC23 models. All models were trained to optimize the objective
\begin{align}
    \mathcal{L}=\lambda_E\sum_i |\hat{E_i}-E_i| +\lambda_F\sum_{i,j}\frac{1}{3N_i} |\hat{F_{ij}}-F_{ij}|^2
\end{align}
where $E_i$ and $\hat{E_i}$ are, respectively, the ground truth and predicted energies of system $i$ with $N_i$ atoms, and
$F_{ij}$ and $\hat{F_{ij}}$ are, respectively, the ground truth and predicted forces for the $j$-th atom in system $i$.
The loss coefficients $\lambda_E$ and $\lambda_F$ are hyperparameters used to trade-off the force and energy losses.
For each model, we used the same model sizes that were originally published, but we tuned the learning rate and the loss coefficients.

Unlike ODAC23 models, which were trained to directly predict the adsorption energy, we use the total DFT energy (with the k-point corrections described in \cref{sec:kpoint_correction}) as the energy target for training our models. To improve training stability and convergence, we apply a linear reference to these energies using the same protocol that was used in the OC22 paper~\cite{Tran2023Electrocatalysis}.
Once trained,
the %
adsorption energy can be calculated by subtracting the energies of the lowest-energy bare MOF and energies of adsorbates from the energy of the combined MOF-adsorbate system:

\begin{equation}\label{eqn:ads_energy}
\begin{aligned}
    \hat{E}_{\text{ads}} =~&\hat{E}_{\text{system}}(r_{\text{system}}) - \hat{E}_{\text{MOF}}(r_{\text{system}}) -
    \sum_{\text{adsorbate}} E_{\text{adsorbate}}(r_{\text{adsorbate}})
\end{aligned}
\end{equation}

where $\hat{E}_{\text{system}}(r_{\text{system}})$ is the predicted energy of the combined MOF-adsorbate system,
$\hat{E}_{\text{MOF}}$ is the energy of the lowest energy bare configuration of the corresponding MOF (as described in \cref{sec:rerelaxations}),
and $E_{\text{adsorbate}}(r_{\text{adsorbate}})$ is the energy of a single adsorbate in the gas phase. The summation is performed over all adsorbates.
In our model evaluation, we compare the adsorption energy predicted by various MLIPs with the adsorption energies derived from DFT calculations.

\setlength{\tabcolsep}{6pt}
\begin{table}[htb!]
\centering
\small
\caption{Overall performance of various models on the S2EF task across all adsorbates. Total Energy MAE (EMAE-Tot) and Adsorption Energy MAE (EMAE) are reported in eV, and Force MAE (FMAE) is reported in \eva. The best result in each column is shown in bold.}
\label{tab:mlip_results_new}
\begin{tabular}{llccc}
\toprule
\textbf{Model} & \textbf{Training Set} & \textbf{EMAE-Tot} & \textbf{EMAE} & \textbf{FMAE} \\
\midrule
EquiformerV2 large~\cite{ODAC23} & ODAC23 & -- & 0.155 & 0.024 \\
MACE~\cite{batatia2024foundationmodelatomisticmaterials} & MPtraj~\cite{deng2023chgnet-39f} & -- & 0.323 & 0.369 \\
MACE-DAC~\cite{mace_dac} & MPtraj~\cite{deng2023chgnet-39f}+GoldDAC & -- & 0.222 & 0.289 \\
\midrule
UMA-Small 1.1~\cite{Wood1985} & ODAC25 subset+others & 0.282 & 0.110 & 0.040 \\
eSEN (Full) & ODAC25-Full & \textbf{0.195} & \textbf{0.077} & \textbf{0.023} \\
eSEN (Filtered) & ODAC25-Filtered & 0.268 & 0.080 & 0.033 \\
\bottomrule
\end{tabular}
\end{table}
\setlength{\tabcolsep}{1.4pt}

\Cref{tab:mlip_results_new} shows S2EF results for various models.
The UMA-Small model~\cite{Wood2025Uma} extends eSEN using mixture of linear experts (MOLE) and was trained on a subset of ODAC25 (\ce{CO2} and \ce{H2O}, without functionalized MOFs) jointly with other datasets.
The two eSEN models~\cite{fu2025learningsmoothexpressiveinteratomic} were trained on the full and filtered ODAC25 datasets, respectively.
All models implement rotational equivariance and constrain forces to be energy-conserving via energy gradients. Since this computation is expensive, models are pre-trained with direct force prediction, then fine-tuned with energy conservation enforced. Hyperparameters are in \Cref{tab:hyperparamaters}.

For comparison, we include EquiformerV2-Large trained on ODAC23 data, the best performing model from that paper, MACE~\cite{Batatia2022mace} a recent foundation model, and MACE-DAC~\cite{mace_dac}, a version of MACE fine-tuned on the GoldDAC dataset. EquiformerV2-Large is based on the EquiformerV2~\cite{liao2023equiformerv2} architecture and contains 153M parameters.

\Cref{tab:mlip_results_new} shows the results of all models on the S2EF task for the MOF + Adsorbate test set of the filtered ODAC25 dataset.
Models trained on ODAC25 significantly outperform prior models. The eSEN models trained on the filtered and full ODAC25 datasets achieve the best overall results with energy MAEs of 0.080 eV and 0.077 eV, and force MAEs of 0.033 \eva~and 0.023 \eva~respectively.
The UMA-Small model, trained on a subset of ODAC25 along with other datasets, shows strong performance with energy MAE of 0.110 eV and a force MAE of 0.040 \eva.
\Cref{tab:mlip_results_by_adsorbate} breaks down the metrics for each adsorbate type.
The eSEN models achieve lower energy and force MAEs on all metrics.
The UMA-Small model is competitive to the eSEN models on  \carbondioxide~and \water, but shows degraded performance for the adsorbates it was not trained on.
These results confirm that ODAC25 provides a strong foundation for training MLIPs that generalize across a wide range of MOF–adsorbate configurations and energies.

\runinhead{Widom insertion and Henry's coefficients.}\label{sec:henrys_coeff}To further evaluate the accuracy of various MLIPs, we calculated Henry's coefficients for \ce{CO2} and \ce{N2} in several MOFs. The Henry coefficient ($K_H$) is the slope of an adsorbate's isotherm in the low-loading limit. The ratio of Henry’s coefficients for two adsorbates gives the adsorption selectivity for the molecular pair in the low loading limit, and this selectivity is often representative of adsorption performance over a wide range of pressures \cite{Walton2015}.

We computed Henry's coefficients using the Widom insertion method~\citep{10.1016/C2009-0-63921-0}; details are provided in the Methods section. Our benchmark dataset consists of several well-known MOF structures (UiO-66, HKUST-1, MOF-5) from the curated experimental dataset published in~\citep{10.1038/s41586-024-07683-8}. Multiple independent experimental isotherms are available for each of these materials \cite{Park2017}. These materials are expected to only involve physisorption, where the rigid framework assumption used in our Widom insertion calculations is likely to be reasonable.

We compare the computational value to experimental Henry's coefficients, which were extracted by fitting the first two data points of the published isotherms to a line passing through the origin. When multiple isotherms are available for the same MOF, gas and temperature, we averaged the Henry coefficient. \Cref{tab:henry_exp} shows the experimental Henry coefficients in mol/kg/Pa and converted into energy space in eV.

\setlength{\tabcolsep}{4pt}
\begin{table*}[htb!]
\centering
\small
\caption{Errors of Henry coefficient prediction for different models on CO$_2$ and N$_2$. We show the mean absolute error between predictions and experimental values in eV, which is proportional to the logarithm of the Henry's constant~\citep{Yu2021}. The best result in each column is shown in bold.}
\label{tab:henry_coefficient_results}
\begin{tabular}{llcc}
\toprule
\textbf{Model} & \textbf{Training Set} & \textbf{CO$_2$} & \textbf{N$_2$} \\
\midrule
UFF & - & 0.0299 & 0.0190 \\
MACE-MP-0b2~\cite{batatia2024foundationmodelatomisticmaterials} & MPtraj \cite{deng2023chgnet-39f} & 0.0566 & 0.0448 \\
MACE-DAC~\cite{mace_dac} & MPtraj~\cite{deng2023chgnet-39f} + GoldDAC & 0.0513 & 0.0698 \\
\midrule
UMA-Small 1.1~\cite{Wood1985} & ODAC25 subset + others & 0.0274 & 0.0631 \\
UMA-Medium 1.1~\cite{Wood1985} & ODAC25 subset + others & \textbf{0.0145} & 0.0286 \\
eSEN & ODAC25-Full & 0.0250 & 0.0433 \\
eSEN & ODAC25-Filtered & 0.0234 & \textbf{0.0136} \\
\bottomrule
\end{tabular}
\end{table*}
\setlength{\tabcolsep}{1.4pt}

\Cref{tab:henry_coefficient_results} shows the results for Henry coefficient predictions across different models for CO$_2$ and N$_2$ adsorbates.
For both \ce{CO2} and \ce{N2}, we see that the best performing model is the one trained on more \ce{CO2} and \ce{N2} data, respectively.
For \ce{CO_2} adsorption, all ODAC25 models outperform the UFF baseline, while the MACE foundation model and the specialized MACE-DAC perform substantially worse. Of the ODAC25 models, the largest, UMA-Medium 1.1, performs best.
For N$_2$ adsorption, the UMA models, which were not trained on N$_2$, as well as the MACE models, do not surpass the UFF baseline, while the eSEN model trained on the filtered ODAC25 data performs best.
Scatter plots of the experimental and predicted values are shown in \cref{fig:suppl-widom}. Methods exist for incorporating MOF flexibility into computation of Henry’s coefficients and adsorption isotherms when force fields are available for the MOF degrees of freedom \cite{Yu2021_flex}, so an interesting future direction will be to adapt these methods for use with MILPs.

Experimental Henry coefficient data in MOFs is scarce and experimental errors are often significant. Comparison of model errors in \cref{tab:henry_coefficient_results} to the spread of experimental Henry coefficients shows that experimental uncertainty accounts for a large portion of model errors. The average spread of experimental Henry coefficients where duplicate data are available at 298 K is 0.016 and 0.009 eV for \ce{CO2} and \ce{N2}, respectively. These spreads approach the MAE of the best performing model for both adsorbates, confirming that experimental accuracy is a limiting factor in further improving models for Henry coefficient prediction.

\runinhead{Adsorption in deformable MOFs.}\label{sec:deform}MOF flexibility can have a non-negligible effect on adsorption energies but is often ignored in simplified calculations using rigid MOFs \cite{Senkovska2023,Agrawal2019,Dundar2017}. ODAC25 allows all MOF atoms to move during relaxation with the adsorbate molecules. Not holding MOFs rigid complicates adsorption energy predictions because models must accurately describe both host-guest interactions and the energy of MOF deformation. A recent study showed that ML force fields achieve mixed results for describing adsorbate-induced MOF deformation \cite{Brabson2025}. We benchmarked the ODAC25 eSEN models using the formalism and dataset of \ce{CO2} and \ce{H2O} adsorption in 59 MOFs presented in that work. \Cref{tab:deform_results} shows the mean absolute errors in adsorption, host-guest interaction, and MOF deformation energies relative to DFT calculations performed in ref. \cite{Brabson2025}. The dataset and code used for this benchmarking are available on the \texttt{fairchem} GitHub repository.

The ODAC25 eSEN models meet or outperform every model tested in ref. \cite{Brabson2025} for predicting the overall adsorption energy. Results using CHGNet \cite{deng2023chgnet-39f}, the best performing force field from ref. \cite{Brabson2025}, are provided for comparison. Results using the ODAC23 direct adsorption energy prediction model (EqV2) are also included but cannot be decomposed into interaction and MOF deformation energies due to its architecture \cite{ODAC23}. Training models on ODAC25 improves their performance on this deformation benchmark, and the eSEN model trained on the filtered set achieves the 0.1 eV $E_{ads}$ goal outlined in ref. \cite{Brabson2025}. The additional data and training on total formation energies introduced in ODAC25 enables our models to outperform the ODAC23 EqV2 model for this task.

Ref. \cite{Brabson2025} cautions that the adsorption energy MAE may be misleading for evaluating models because the errors are typically of the same magnitude as the adsorption energies. Predictions from a good model should correlate well with the DFT ground truth, so $R^2$ is a useful alternative metric. The $R^2$ scores are –1.264 and 0.542 for the full and filtered eSEN models, respectively. The negative score indicates that the model predictions are worse than simply guessing the mean of the DFT adsorption energies. eSEN-Filtered outperforms every model tested in ref. \cite{Brabson2025} using $R^2$.

A related question is whether a model can determine whether or not a MOF undergoes significant deformation during adsorption. Taking a MOF deformation energy of 0.05 eV as the boundary between ``Negligible" and ``Significant" MOF deformation, \cref{fig:confusion} shows that both eSEN models struggle to identify MOFs which undergo significant deformation. Every model tested in ref. \cite{Brabson2025} shows similar behavior, with eSEN-Filtered again outperforming all models for this task. Despite further room for improvement in adsorption energy MAE, $R^2$, and deformation classification, training more advanced ML architectures on ODAC25 appears promising for developing models capable of accurately predicting adsorption in deformable MOFs.

\setcounter{section}{2}
\section{Discussion}
\label{sec:conclusion}

The ODAC25 dataset represents a step forward in computational sorbent discovery for DAC, addressing critical gaps in DFT accuracy and chemical diversity. ODAC25 improves MOF validation, k-point sampling, and treats adsorbate-induced MOF deformation using a physically relevant approach. It introduces \ce{N2} and \ce{O2} adsorbates, functionalized and synthetically generated MOFs, and high-energy GCMC configurations, providing a more comprehensive foundation for modeling competitive adsorption and multicomponent interactions.

By leveraging advanced data generation methods, ODAC25 captures a diverse range of adsorption phenomena, from single-molecule chemisorption to complex multicomponent configurations. The integration of Grand Canonical Monte Carlo (GCMC) data, paired with single-point DFT calculations, bridges the gap between atomistic modeling and process-level predictions, enhancing the dataset's relevance to both high-throughput screening and machine learning model training.

The updated machine learning force fields described above, trained on the expanded dataset, demonstrate improved performance on both ODAC23 test sets and new functionalized MOFs, highlighting the value of the extended data. These models advance the field by better capturing complex interactions and enabling accurate predictions across a wider range of sorbent chemistries and conditions. By making the dataset, models, and tools publicly available, we aim to empower the community to accelerate the development of scalable and energy-efficient sorbents for carbon dioxide removal.

\setcounter{section}{3}
\section{Methods}
\label{sec:methods}

\subsection*{MOF functionalization}
We developed our MOF functionalization process in Python, building on a previous MOF point defect generator \cite{Yu_Jamdade_Yu_Cai_Sholl_2023}. An advantage of our approach is that it eliminates the need for user-specified substructures and instead automatically functionalizes MOFs using predefined methods. In addition, the package supports user-defined molecules for OMS functionalization.

We functionalized 110 pristine and 65 defective MOFs with PLD $\geq$ 10 Å to allow enough space for amine groups and \ce{CO2} adsorption. During the linker functionalization process, we used MOFid \cite{MOFid_2019} to separate metal centers and linkers. The separated linkers were then compared to the linker candidates in \cref{tab_linkers}. If a match was found, the corresponding modification was applied to the linker by adding amine functional groups. Original linkers were functionalized at all possible concentrations, with concentration defined as the number of sites modified relative to the total available sites. For OMS functionalization, we used the OMS detection algorithm developed in the CoRE MOF 2019 database~\cite{Core_MOF_2019}. If an OMS was detected in a MOF structure, diamines were grafted at all possible concentrations in the unit cell. All ten diamines in \cref{tab_diamines} were used. In each case the more primary amine was appended to the OMS. The orientations of the diamines were pre-optimized to avoid overlapping atoms, then structures were DFT relaxed. This generated a total of 7,163 distinct DFT-relaxed amine-functionalized MOF structures.
The PLDs of all functionalized MOFs were calculated after DFT relaxation, and only structures with PLD $\geq$ 3.3 Å were subsequently used for adsorbate placement and DFT relaxation.

We used DFT to relax adsorbed \ce{CO2} and \ce{H2O} in all functionalized MOFs. We first used the ODAC23 strategy for the placement of [\ce{CO2}], [\ce{H2O}], [1\ce{CO2}+1\ce{H2O}] and [1\ce{CO2}+2\ce{H2O}] in all functionalized MOFs \cite{ODAC23}. A limitation of this approach is that it relies on a classical FF that does not allow configurations corresponding to reactions between molecules and amine groups. If an energy barrier exists between these states and configurations involving reactions, DFT relaxation from the initial state will not find the physically interesting latter state.

To address this limitation, we performed additional calculations specifically aimed at generating initial configurations similar to those known from \ce{CO2} reactions with amines. We investigated two \ce{CO2} reactive capture mechanisms involving amines. The first mechanism is the cooperative formation of ammonium carbamate~\cite{McDonald_Long_2015,Vlaisavljevich_Smit_2015,Siegelman2017,Milner_Long_2017,Forse_long_2018,Zhu_long_2024,Kim_Long_2020}. \ce{CO2} inserts into each metal amine bond, forming ion-paired ammonium carbamate chains (illustrated in \cref{fig_mechanism}a). This mechanism has been studied in detail previously in a DFT climbing-image nudged elastic band (CI-NEB) study in mmen-$\mathrm{Mg_2(dobpdc)}$ \cite{Lin_Jiang_2023}.

We also generated initial states involving carbamic acid as shown in \cref{fig_mechanism}b. Compared with 2:1 amine:\ce{CO2} stoichiometry for the mechanism generating ammonium carbamate, this represents a 1:1 amine:\ce{CO2} stoichiometry. Carbamic acids have been experimentally identified in \ce{CO2} capture using post-synthesized OMS amine functionalized MOFs \cite{Lee_Hong_2014,Forse_long_2018}. For MOFs with linker functionalization, \ce{CO2} reacts with the amine groups on the linker to form the carbamic acid (\cref{fig_mechanism}c). This mechanism has been observed experimentally using NMR in linker-functionalized IRMOF-74-III-$\mathrm{CH_2NH_2}$ \cite{linker_Fracaroli_Yaghi_2014} and IRMOF-74-III-$\mathrm{(CH_2NH_2)_2}$ \cite{linker_Popp_yaghi_2017}.

For linker functionalized MOFs, one random functionalized amine group (\ce{-NH2}) was selected and replaced with carbamic acid group (-NHCOOH). For OMS functionalization, one random grafted diamine was selected and replaced with corresponding ammonium (\ce{NR3H+}) and carbamate (\ce{NR2COO^{–}}), or carbamic acid \ce{NR2COOH} if the outer amine was primary or secondary, respectively. For each functionalized MOF with 1 \ce{CO2} placed using one of the reactive placements outlined above, we also created structures with 1 and 2 \ce{H2O} to generate functionalized structures with [1\ce{CO2}+1\ce{H2O}] and [1\ce{CO2}+2\ce{H2O}]. In total, we successfully generated 52,756 adsorbate placements in 3,920 functionalized MOFs. Automated generation of functionalized MOFs is challenging, and some of our structures may not be experimentally relevant or accessible, but all of the included structures correspond to converged DFT calculations and provide additional diversity to the dataset.

\subsection*{High-energy configuration GCMC simulation settings}
GCMC simulations were run at 300 K for pressures of 5, 10, 20, and 50 kPa. To explore mixed-component adsorption behavior, we performed simulations with pure \ce{CO2} and pure \ce{H2O} and mixtures with \ce{CO2}-\ce{H2O} gas phase molar ratios of 1:1, 1:5, 1:10, and 1:20. Because adsorption in MOFs is selective (typically for \ce{H2O} relative to \ce{CO2}) this approach generates a wide range of adsorbed compositions. Each simulation was run for 500,000 cycles.

\subsection*{Synthetic MOF generation}
An autoregressive transformer model samples MOF specifications in the form of a sequence of the topology, the identities of the metal clusters, and the identities of the ligands. We assemble these specifications into atomistic structures using Pormake~\citep{10.1021/acsami.1c02471}. The settings used to construct these MOF structures are included in the supplementary material.

The generated structures were initially optimized with the UFF4MOF force field~\citep{UFF4MOF2}. This relaxation protocol includes a phase of MD simulation at room temperature, and we rejected structures in which the unit cell explodes or collapses. MOFChecker v0.9.6~\citep{D5DD00109A} was used to screen the structures for overlapping atoms and improper metal coordination as described in \cref{subsubsec:validation}. In addition to rejecting problematic structures flagged by MOFChecker, we excluded structures with more than 250 atoms in the unit cell, PLDs less than 3.64 Å or more than 20 Å, and structures containing lanthanides or actinides.

Of the remaining 4,000 structures, we selected the 460 that improve the diversity of ODAC25 the most. We use farthest point sampling~\citep{10.1016/0304-3975(85)90224-5} in a space that combines several features that characterize MOFs, including geometric properties related to porosity and surface area, autocorrelation functions~\citep{10.1021/acs.jpca.7b08750}, as well as the heat of adsorption of CO$_2$. We computed these features with mofdscribe~\citep{Jablonka_2022}, RASPA~\citep{RASPA}, and Zeo++~\citep{Zeo++}.%

The selected structures were relaxed with DFT. For the limited number of structures for which this relaxation failed, the structures were removed and replaced with new candidates, again using farthest point sampling approach.

\subsection*{Widom insertions}
The Widom insertion method involves randomly inserting test molecules into the rigid MOF structure and computing their interaction energy. We approximated the MOF as being rigid in these calculations, so the MOF-adsorbate interaction energy is given by Eq. \ref{eqn:ads_energy}. The Henry's coefficient is then calculated as
$$K_H = \frac{\beta}{V} \langle \exp(-\beta \hat E_{\text{int}}) \rangle$$
where $\beta = 1/k_BT$, $V$ is the system volume, and the angle brackets denote an expectation value over uniformly distributed insertion positions and orientations \cite{Yu2021}. Because the Henry’s coefficient averages over the entire pore structure, it represents an example of using force fields to compute a property that cannot be assessed from a small collection of DFT calculations.

As a baseline, we perform these calculations using the Universal Force Field (UFF)~\citep{10.1021/ja00051a040} as implemented in the RASPA2 simulation package~\citep{RASPA} with the Automated Interactive Infrastructure and Database for Computational Science (AiiDA) framework~\citep{10.1016/j.commatsci.2015.09.013}. The baseline uses the Transferable Potentials for Phase Equilibria (TraPPE) force field~\citep{10.1002/aic.690470719} for the \ce{CO2} and \ce{N2} molecules. To evaluate the MLIPs introduced in this paper, we employ a Python package\footnote{Released at \url{https://github.com/Cusp-AI/widom}} based on DAC-SIM \cite{mace_dac} that can perform Widom insertion calculations with any Atomic Simulation Environment (ASE) \citep{ase-paper} calculator. This allows us to directly compare the performance of different MLIPs against the UFF baseline.

\setcounter{section}{4}
\section*{Acknowledgments}
\label{sec:acknowledgment}

We acknowledge Jin, Garcia, and Smit~\cite{jin2025correspondence} for their correspondence on the ODAC23 dataset and for helpful discussions. We also appreciate detailed comments on the initial preprint from Andrew Rosen that led to several improvements, including a more nuanced discussion of the effects of spin on adsorption. DSS acknowledges fundings from the Oak Ridge National Laboratory LDRD Program.

\section*{Author Contributions}
AS, LMB, and XY are co-lead authors on this work. AS, ZU, AJM, and DSS conceived the study. LMB performed screening of MOFs for chemical reasonableness, MOF re-relaxations, and the deformation analysis. AS performed DFT relaxations, k-point convergence testing and GCMC configurations from experimental MOFs. SC and LMB carried out \ce{N2}/\ce{O2} adsorbate placements. XY and LMB generated diamine functionalized structures and the corresponding adsorbate placements. EM, PdH, SL, JB, and MW generated synthetic MOFs and high-energy configurations from GCMC, and PdH led the Widom insertion study. AS, KA, JRK, CLZ, and ZU were responsible for training and fine-tuning the ML models. AS, JRK, MW, CLZ, ZU, AJM, and DSS provided supervision for the study. All coauthors contributed to the text.

\section*{Competing Interests}
The authors declare no competing interests.

\section*{Data Availability Statement}
The ODAC25 dataset and model checkpoints are available at \newline https://huggingface.co/facebook/ODAC25.

\section*{Code Availability Statement}
The Widom insertion code is available at https://github.com/cusp-ai-oss/widom.
The MOF functionalization code is available at

\clearpage
\newpage
\bibliographystyle{IEEEtran}
\bibliography{paper}

\clearpage
\newpage

\beginappendix

\setcounter{figure}{0}
\setcounter{table}{0}
\renewcommand{\thefigure}{S\arabic{figure}}
\renewcommand{\thetable}{S\arabic{table}}
\pagenumbering{arabic}
\setcounter{page}{1}
\renewcommand{\thepage}{S\arabic{page}}

\paragraph{\textbf{Data Availability Statement}} The full ODAC25 dataset and all the trained ML models are publicly available at \url{https://huggingface.co/facebook/ODAC25}.

\section{MOFChecker Analysis}
\begin{table*}[ht!]
    \centering
    \renewcommand{\arraystretch}{1.0}
    \setlength{\tabcolsep}{5pt}
    \caption{Error rates from MOFChecker v0.9.6 checks split by MOF types}
    \resizebox{\textwidth}{!}{
    \begin{tabular}{lcccc}
        \toprule
        Check & \ ODAC23 pristine [\%] & \ ODAC23 defective [\%] & \ ODAC25 functionalized [\%] & \ All MOFs [\%] \\
        \midrule
        \texttt{has\_metal} & \ 0.02 & \ 0 & \ 0 & \ $<0.01$ \\
        \texttt{has\_carbon} & \ 4.4 & \ 6.6 & \ 0 & \ 4.1 \\
        \texttt{has\_hydrogen} & \ 10.4 & \ 0 & \ 0 & \ 3.5 \\
        \texttt{has\_atomic\_overlaps} & \ 0 & \ 0 & \ 0 & \ 0 \\
        \texttt{has\_overcoordinated\_carbon} & \ 0.5 & \ 0.03 & \ 0 & \ 0.2 \\
        \texttt{has\_overcoordinated\_nitrogen} & \ 0 & \ 0 & \ 0 & \ 0 \\
        \texttt{has\_overcoordinated\_hydrogen} & \ 0 & \ 0 & \ 0 & \ 0 \\
        \texttt{has\_undercoordinated\_carbon} & \ 10.9 & \ 23.6 & \ 13.3 & \ 16.7 \\
        \texttt{has\_undercoordinated\_nitrogen} & \ 5.9 & \ 11.0 & \ 0.7 & \ 6.6 \\
        \texttt{has\_undercoordinated\_rare\_earth} & \ 0.02 & \ 0.06 & \ 0 & \ 0.03 \\
        \texttt{has\_undercoordinated\_alkali(ne)} & \ 0.5 & \ 0.7 & \ 0 & \ 0.4 \\
        \texttt{has\_lone\_molecule} & \ 10.3 & \ 32.2 & \ 23.4 & \ 22.6 \\
        \texttt{has\_high\_charges} & \ 1.0 & \ 0.8 & \ 0 & \ 0.7 \\
        \texttt{has\_suspicious\_terminal\_oxo} & \ 0.2 & \ 1.4 & \ 1.4 & \ 1.0 \\
        \texttt{has\_geometrically\_exposed\_metal} & \ 6.5 & \ 8.5 & \ 2.9 & \ 6.4 \\
        \bottomrule

    \end{tabular}
    }
    \label{tab:mofcheck}
    \vspace{-10pt}
\end{table*}

\FloatBarrier

\section{K-point Corrections}
\begin{figure*}[h]
    \centering
    \tiny
    \includegraphics[width=\textwidth]{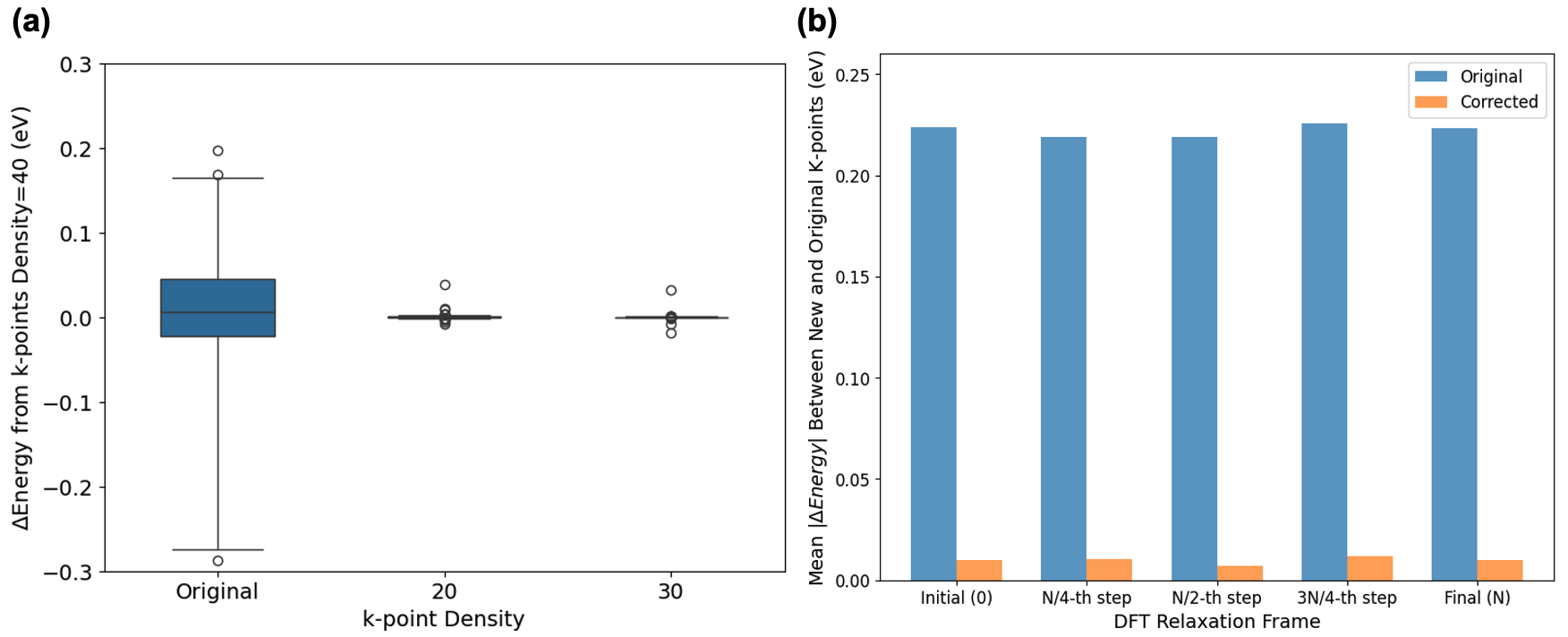}
    \caption{
    Approximating higher k-point calculations. (a) Distribution of energy differences from reference calculations with k-point density $K=40$ \AA{}. (b) Mean absolute energy errors across relaxation frames before and after energy correction. The mean error reduces significantly after correction to $\sim 0.01$ eV.
    }
    \label{fig:convergence_errors_correction}
\end{figure*}

\clearpage
\FloatBarrier
\section{Density Functional Theory Settings}
\begin{table*}[!h]
    \caption{Example INCAR file showing the VASP settings used in creating the ODAC25 dataset. All DFT calculations used VASP version 6.3 \cite{vasp1,vasp2,vasp3} with the PBE exchange-correlation functional and D3 van der Waals correction (IVDW = 12). VASP 5.4 PBE pseudopotentials were employed with a plane-wave energy cutoff of 600 eV and electronic convergence of 1e-5 eV. Structural relaxations used the conjugate gradient algorithm with forces converged to 0.05 \eva~or a maximum of 2000 ionic steps. Bare MOFs were relaxations allowed  relaxation of lattice parameters (ISIF = 3), while MOF+adsorbate systems used fixed lattice parameters (ISIF = 2) to maintain consistent unit cells for binding energy calculations. Gaussian smearing ($\sigma = 0.2$ eV) was applied with symmetry disabled to allow proper adsorbate-framework relaxation.}
    \label{tab:dft_settings}
      \centering
\scalebox{0.8}{
\begin{tabular}{lc}
\toprule
Parameter & Value\\ \midrule
 ENCUT & 600.000000\\
 POTIM & 0.010000\\
 SIGMA & 0.200000\\
 EDIFF & 1.00e-05\\
 EDIFFG & -5.00e-02\\
 ALGO & NORMAL\\
 GGA & PE\\
 PREC & Accurate\\
 IBRION & 2\\
 ISIF & 2\\
 ISMEAR & 0\\
 ISPIN & 2\\
 ISTART & 0\\
 ISYM & 0\\
 MAXMIX & 40\\
 NELM & 120\\
 NELMIN & 2\\
 NSW & 2000\\
 NWRITE & 2\\
 IVDW & 12\\
 NCORE & 4\\
 LCHARG & .FALSE.\\
 LDIAG & .TRUE.\\
 LPLANE & .TRUE.\\
 LWAVE & .FALSE.\\
 LREAL & Auto\\
\bottomrule
\end{tabular}}
\end{table*}

\clearpage
\FloatBarrier
\section{\ce{CO2} and \ce{H2O} Adsorption Energy Distributions}
\begin{figure}[ht!]
    \centering
    \includegraphics[width=\textwidth, height=\textheight, keepaspectratio]{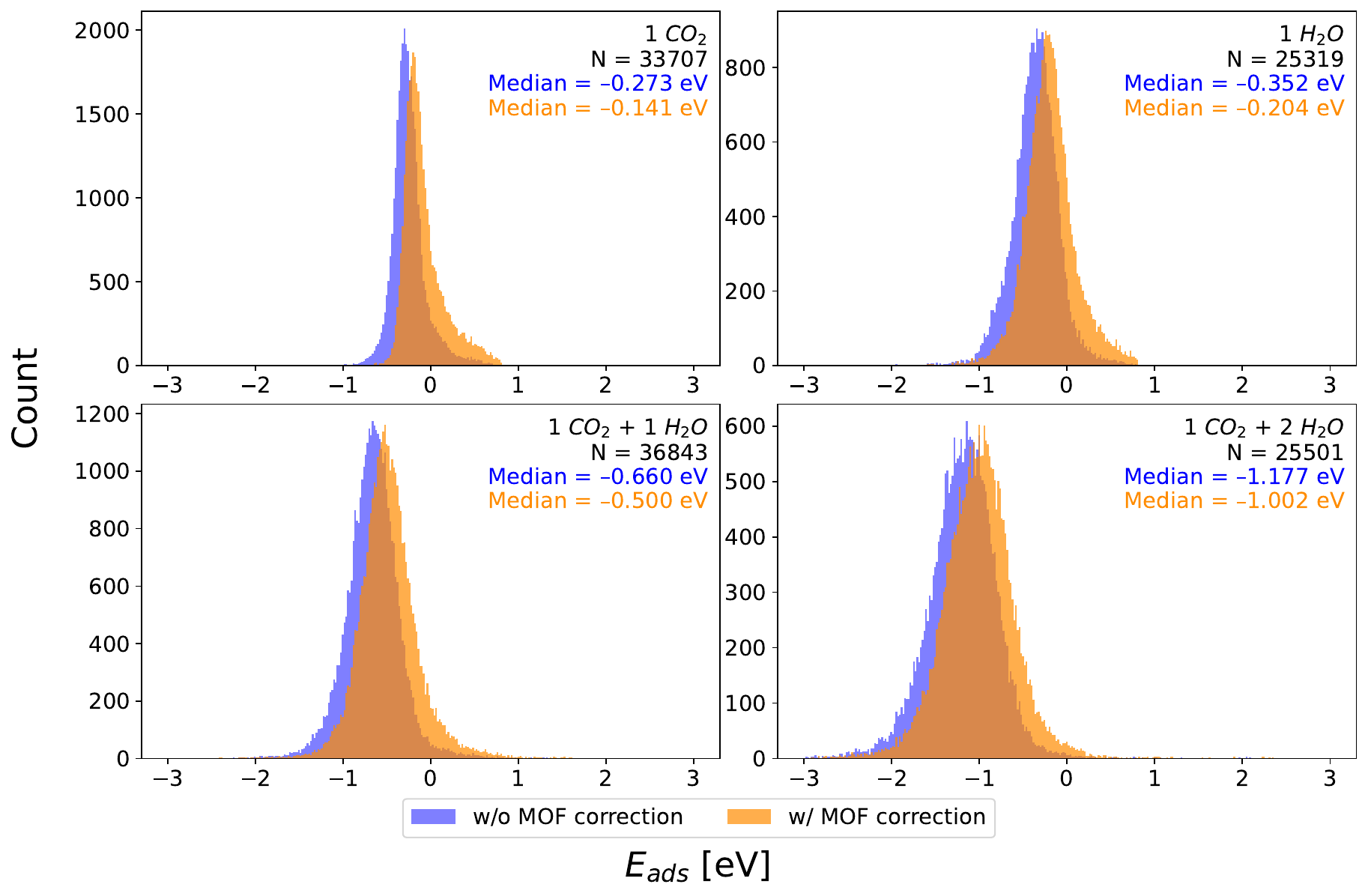}
    \caption{Distribution of \ce{CO2} and \ce{H2O} adsorption energies in pristine and defective ODAC23 MOFs featuring empty MOF re-relaxations and split by adsorbate type(s). Blue histograms are for adsorption energies presented in ODAC23 without MOF re-relaxations; orange histograms are adsorption energies using the global empty MOF reference energy from all ODAC25 configurations with a given MOF.}
    \label{fig:odac23_rerelax_dist}
\end{figure}

\begin{figure}[ht!]
    \centering
    \includegraphics[width=\textwidth, keepaspectratio]{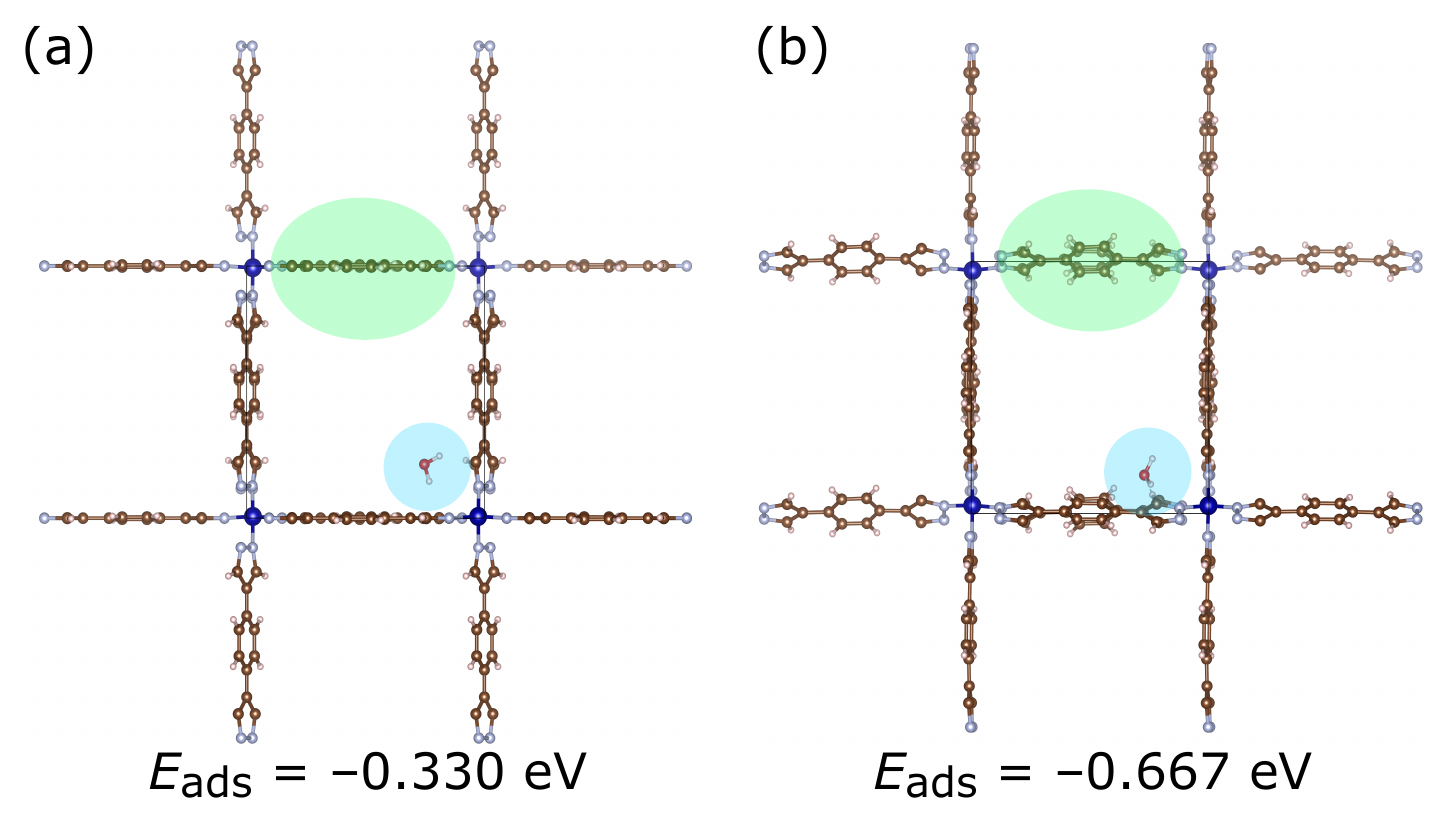}
    \caption{Water adsorption in the MOF with CSD code COJHIT when the MOF is assumed (a) rigid and (b) free to deform. Co, C, N, O, and H atoms are shown in blue, brown, silver, red, and white, respectively.}
    \label{fig:COJHIT_rigid_vs_flex}
\end{figure}

\clearpage
\FloatBarrier
\section{\ce{N2} Adsorption}
\begin{figure*}[ht!]
    \centering
    \includegraphics[width=\textwidth, height=\textheight, keepaspectratio]{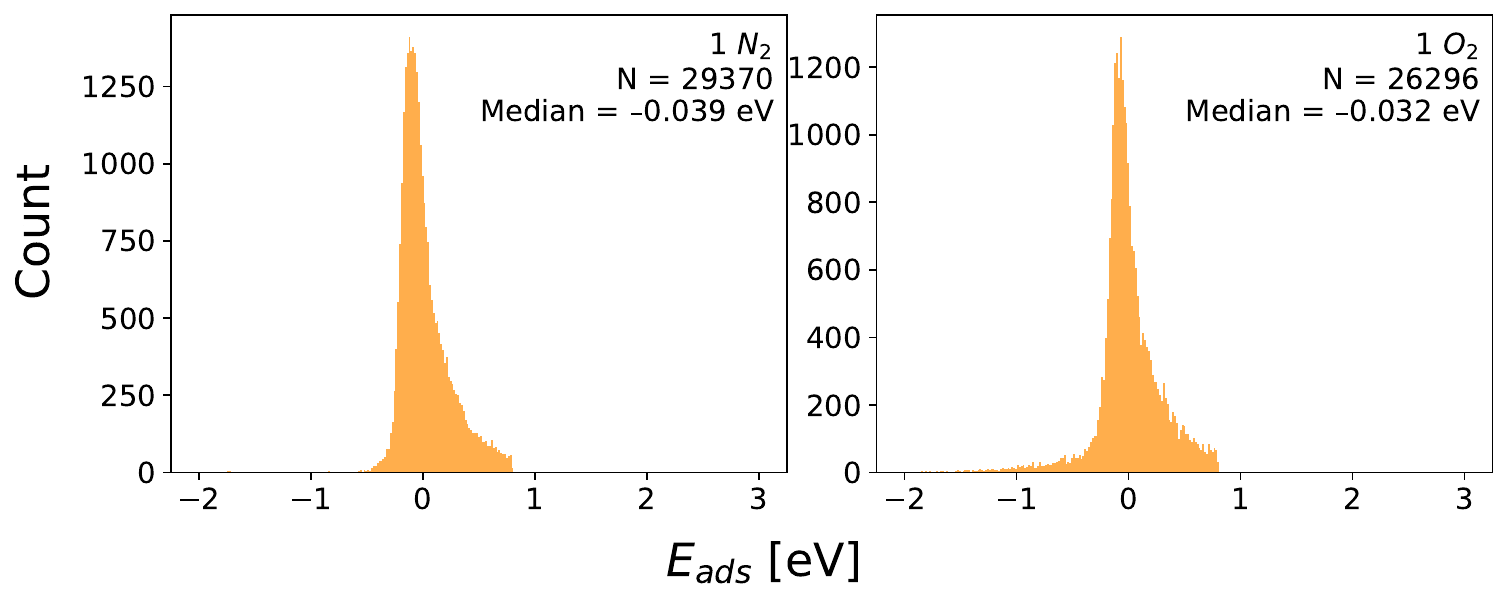}
    \caption{Distribution of \ce{N2} and \ce{O2} adsorption energies in pristine and defective MOFs split by adsorbate type.}
    \label{fig:odac23_n2o2_dist}
\end{figure*}

\begin{figure*}[ht!]
    \centering
    \includegraphics[width=0.7\textwidth, height=\textheight, keepaspectratio]{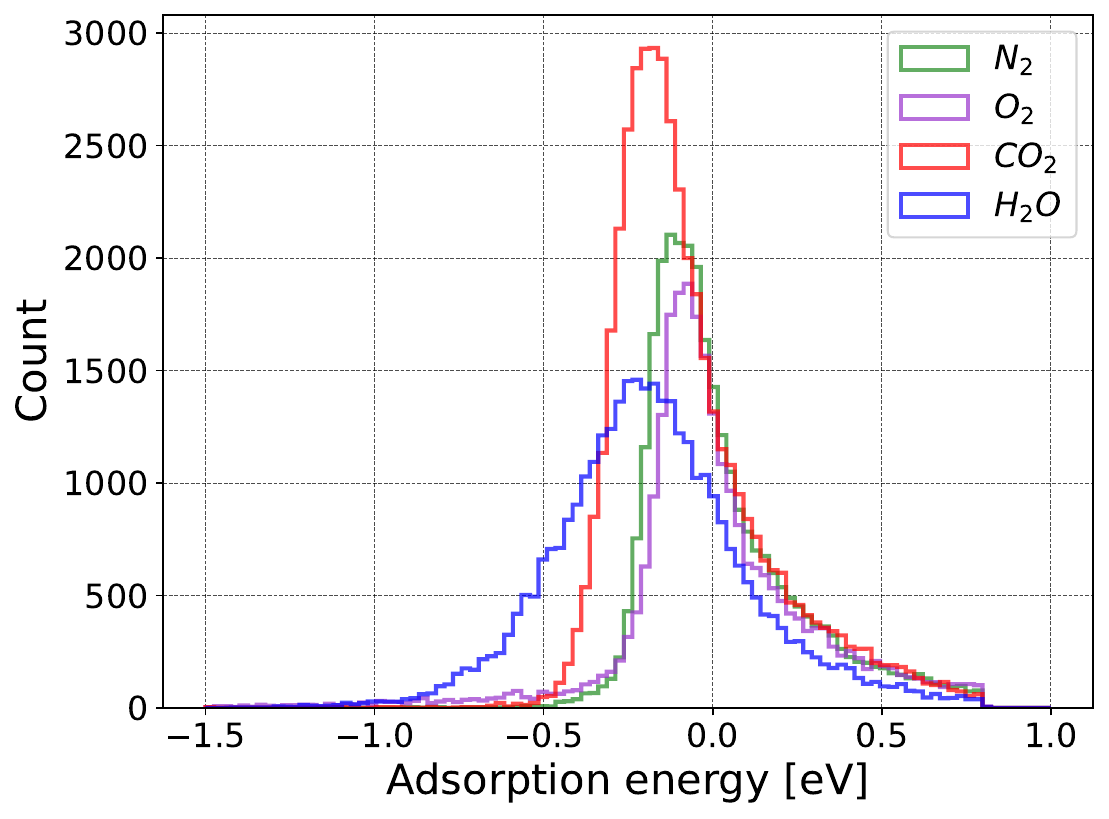}
    \caption{Histogram of \ce{CO2}, \ce{H2O}, \ce{N2}, and \ce{O2} adsorption energies in ODAC23 MOFs split by adsorbate.}
    \label{fig:n2o2_hist}
\end{figure*}

\begin{figure*}[ht!]
    \centering
    \includegraphics[width=0.7\textwidth, height=\textheight, keepaspectratio]{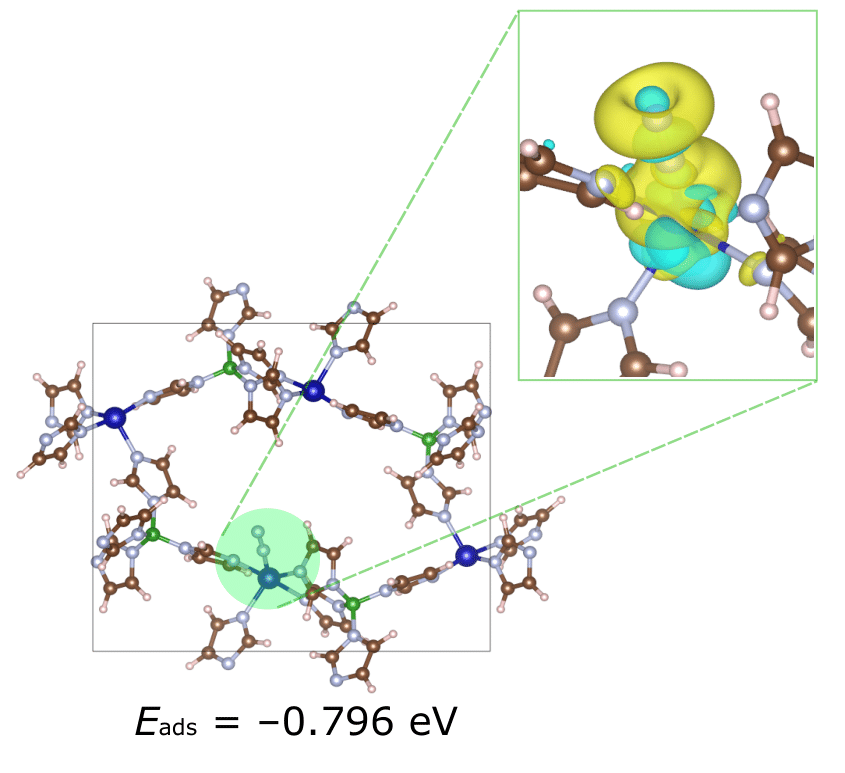}
    \caption{\ce{N2} chemisorption in MOF with CSD code DEJRUH with $E_{ads}$ = –0.796 eV. The inset shows isosurfaces of zero net charge transfer during chemisorption computed using Bader charge analysis. Yellow and cyan volumes show electron acceptors and donors, respectively. Co, B, C, N, and H atoms are shown in blue, green, brown, silver, and white, respectively.}
    \label{fig:n2_chemisorption}
\end{figure*}

\clearpage
\FloatBarrier
\section{MOF Functionalization}
\begin{table}[h]
\centering
\caption{Linkers used to generate functionalized MOFs via linker functionalization.
}\label{tab_linkers}%
\footnotesize
\begin{tabular}{@{} >{\centering\arraybackslash}m{3cm}  %
                        >{\centering\arraybackslash}m{3cm}
                        >{\centering\arraybackslash}m{3cm}
                        >{\centering\arraybackslash}m{3cm} @{}}
\toprule
Linker Structure  & Functionalized Linker Structure & Reference\\
\midrule
\includegraphics[scale=0.2]{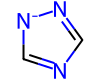}    & \includegraphics[scale=0.2]{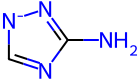}  & \cite{funced_linker_Lin_Chen_2012} \\%
\includegraphics[scale=0.2]{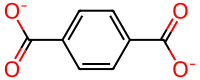}   & \includegraphics[scale=0.2]{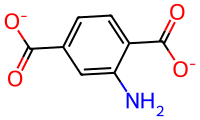} & \cite{funced_linker_Chen_Jiang_2005} \\%
\includegraphics[scale=0.2]{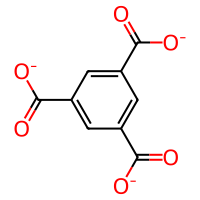}   & \includegraphics[scale=0.2]{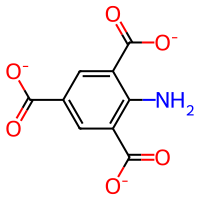} & \cite{funced_linker_Chen_Ng_2023} \\%
\multirow{2}{*}{\includegraphics[scale=0.2]{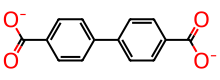}}    & \includegraphics[scale=0.2]{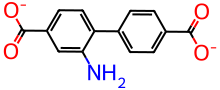}  & \cite{funced_linker_Deshpande_Telfer_2010}\\%
& \includegraphics[scale=0.2]{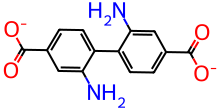}  & \cite{funced_linker_Diamantis_Lazarides_2018}\\%
\includegraphics[scale=0.2]{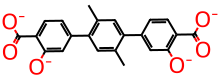}    & \includegraphics[scale=0.2]{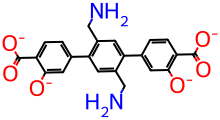}  & \cite{linker_Popp_yaghi_2017}\\%
\includegraphics[scale=0.2]{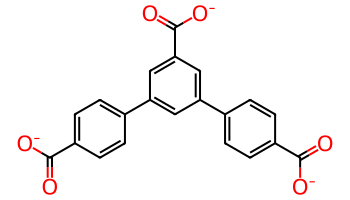}  & \includegraphics[scale=0.2]{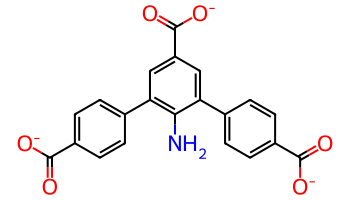} & \cite{funced_linker_Fan_Wang_2016} \\ %
\bottomrule
\end{tabular}
\end{table}

\FloatBarrier
\begin{table}[htb!]
\centering
\caption{
Diamine structures used for open metal site functionalization, spanning primary (1°), secondary (2°), and tertiary (3°) amine classifications with their chemical structures and abbreviations.
}\label{tab_diamines}
\begin{threeparttable}
\footnotesize
\begin{tabular}{@{} >{\centering\arraybackslash}m{6cm}  %
                        >{\centering\arraybackslash}m{4cm}
                        >{\centering\arraybackslash}m{2cm}
                        >{\centering\arraybackslash}m{2.5cm} @{}}
    \toprule
    Diamine Name  & Diamine Structure & Abbreviation & Classification\\
    \midrule
    Ethylenediamine  & \includegraphics[scale=0.25]{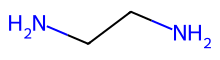}  & en & 1°/1° \\
    N-Methylethylenediamine  & \includegraphics[scale=0.25]{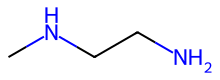}  & nmen & 1°/2° \\
    N-Ethylethylenediamine  & \includegraphics[scale=0.25]{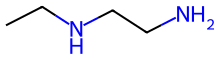}  & een & 1°/2° \\
    N-Isopropylethylenediamine  & \includegraphics[scale=0.25]{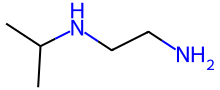}  & ipen & 1°/2° \\
    N,N-Dimethylethylenediamine  & \includegraphics[scale=0.25]{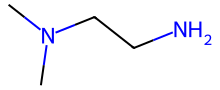}  & dmen & 1°/3° \\
    N,N-Diethylethylenediamine  & \includegraphics[scale=0.25]{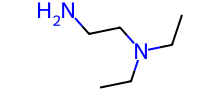}  & deen & 1°/3° \\
    Dimethylethylenediamine  & \includegraphics[scale=0.25]{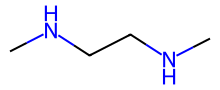}  & mmen & 2°/2° \\
    N,N'-Diethylethylenediamine  & \includegraphics[scale=0.25]{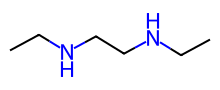}  & eeen & 2°/2° \\
    N,N,N'-Trimethylethylenediamine  & \includegraphics[scale=0.25]{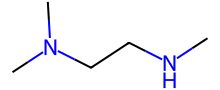}  & mden & 2°/3° \\
    N,N,N',N'-Tetramethylethylenediamine  & \includegraphics[scale=0.25]{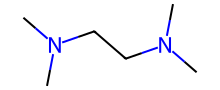}  & tmen & 3°/3° \\
    \bottomrule
\end{tabular}
\begin{tablenotes}
    \item[] 1°: Primary, 2°: Secondary, 3°: Tertiary
\end{tablenotes}
\end{threeparttable}
\end{table}

\FloatBarrier
\begin{figure*}[ht!]
    \centering
    \includegraphics[width=\textwidth, height=\textheight, keepaspectratio]{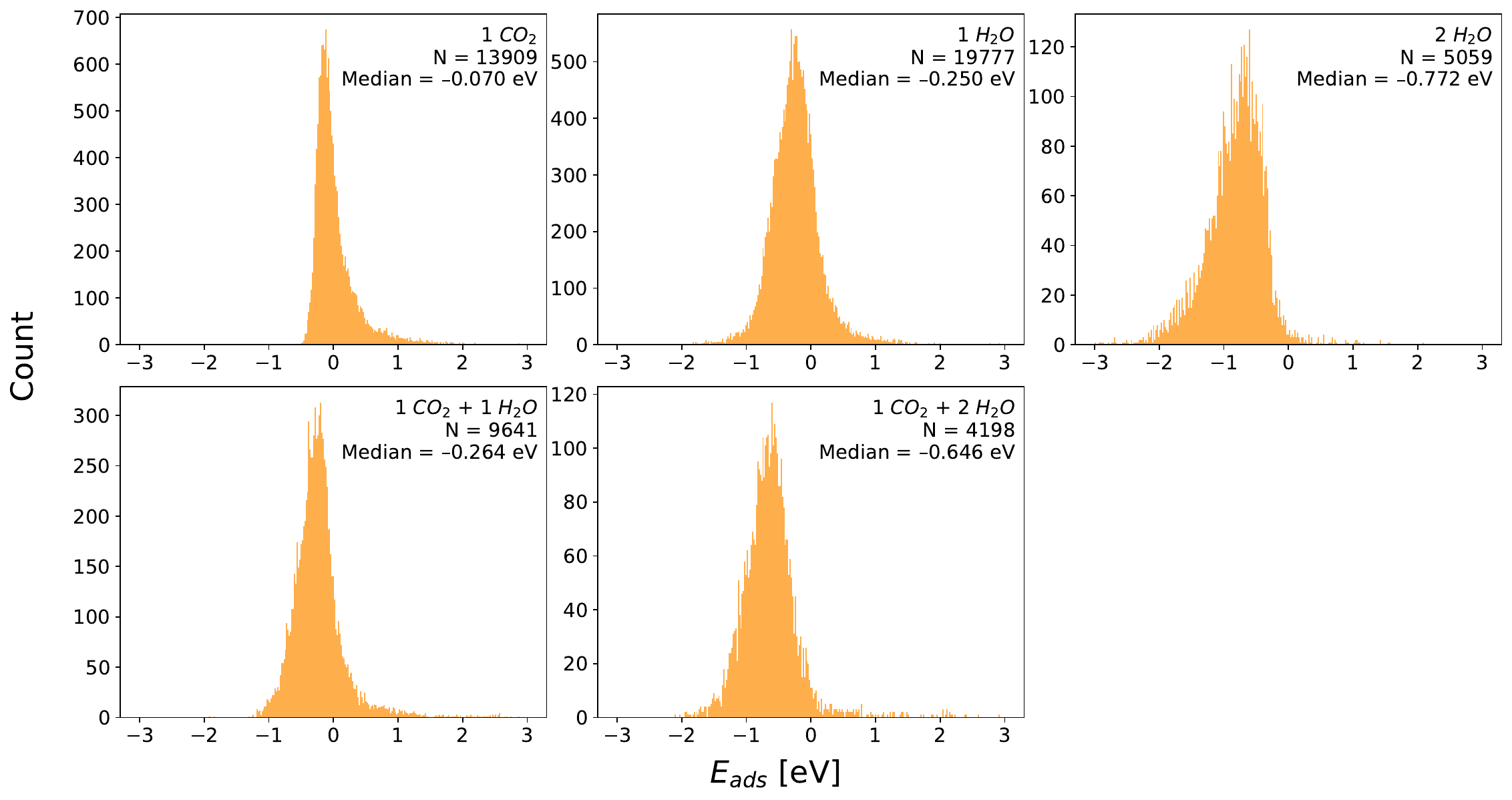}
    \caption{Distribution of adsorption energies in functionalized MOFs split by adsorbate type(s).}
    \label{fig:odac25_func_dist}
\end{figure*}

\begin{figure*}[htb!]
    \centering
    \includegraphics[width=0.7\textwidth, height=\textheight, keepaspectratio]{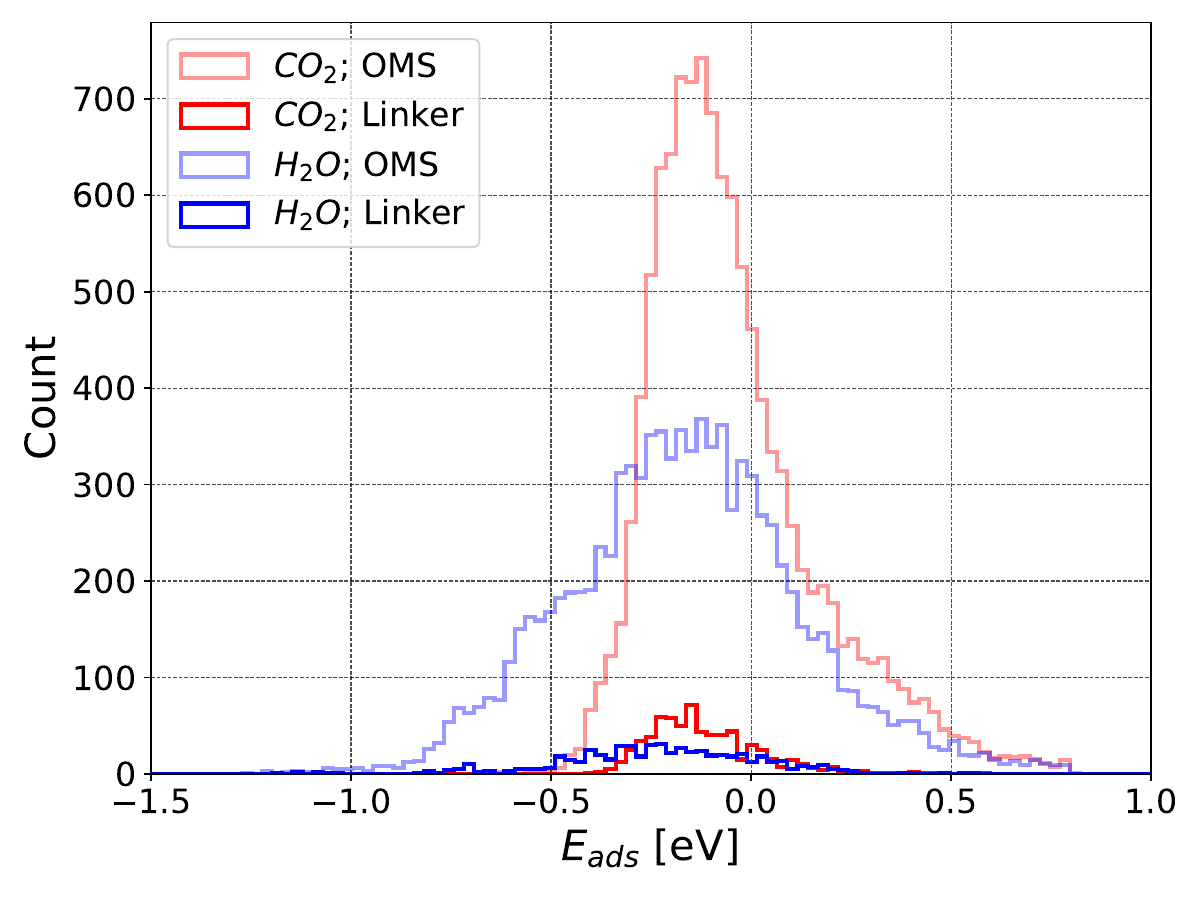}
    \caption{Histogram of \ce{CO2} and \ce{H2O} adsorption energies in functionalized MOFs split by adsorbate and functionalization type.}
    \label{fig:func_type_histogram}
\end{figure*}

\begin{figure*}[htb!]
    \centering
    \includegraphics[width=0.7\textwidth, height=\textheight, keepaspectratio]{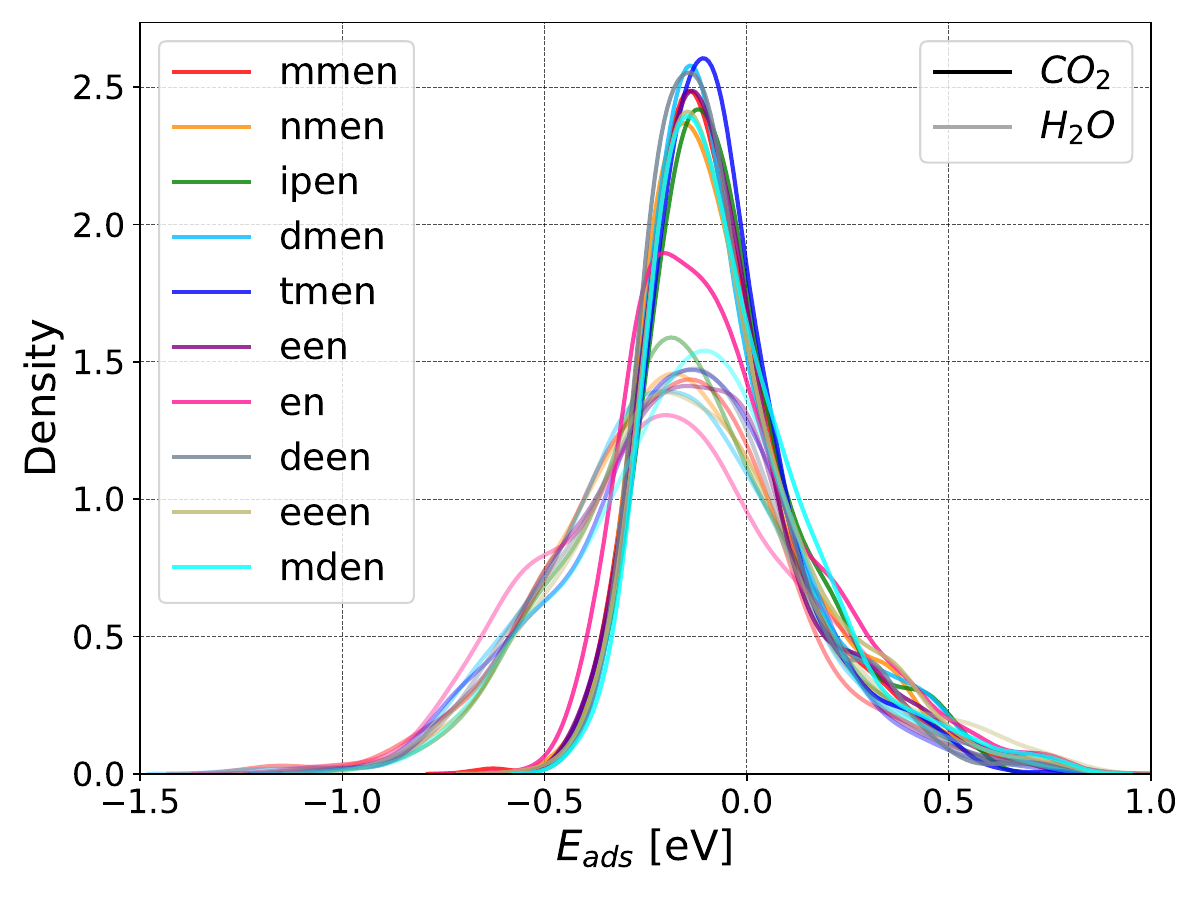}
    \caption{Kernel density estimation for \ce{CO2} and \ce{H2O} adsorption energies in functionalized MOFs split by diamine used for functionalization.}
    \label{fig:diamine_kde}
\end{figure*}

\begin{figure*}[htb!]
    \centering
    \includegraphics[width=\textwidth, height=\textheight, keepaspectratio]{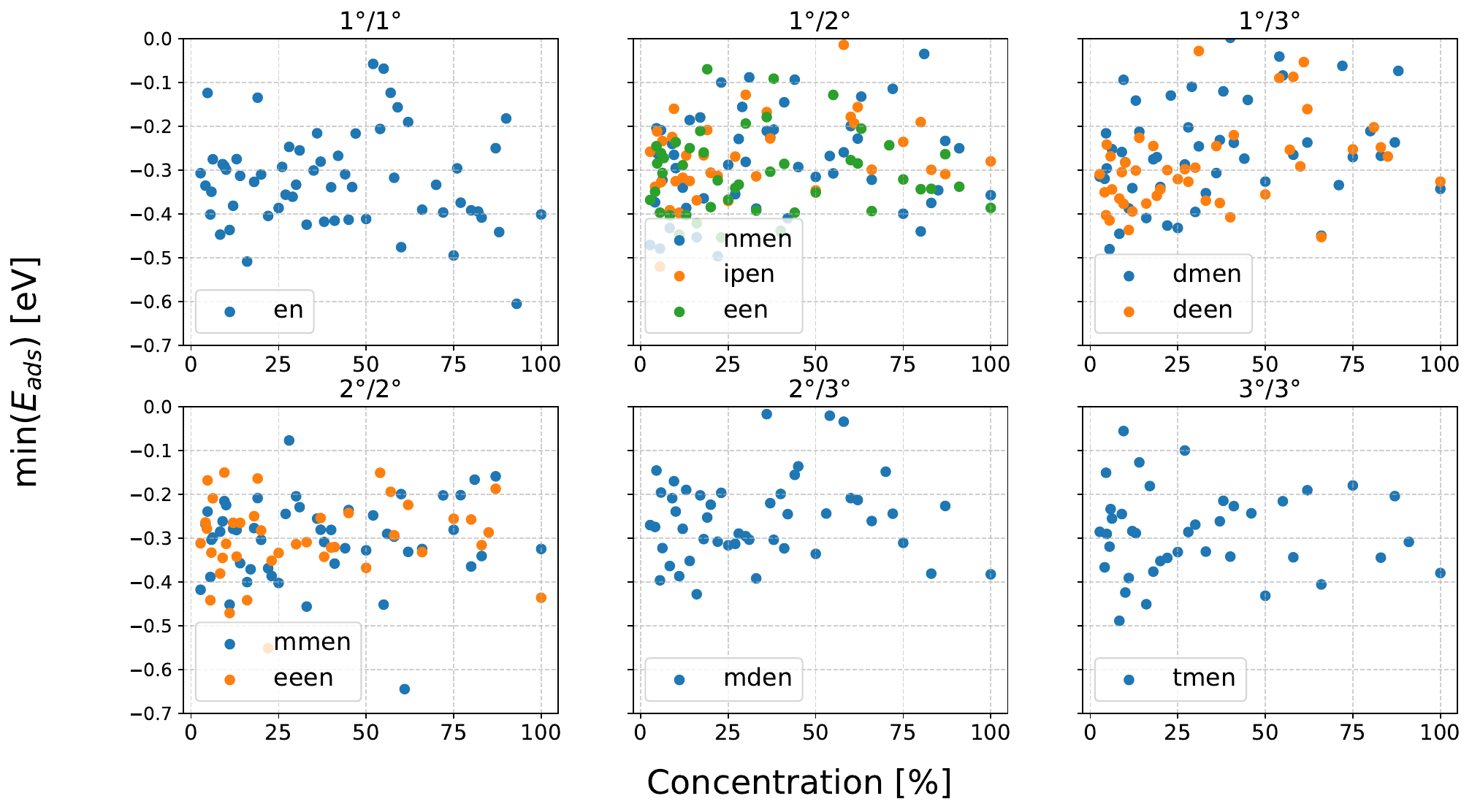}
    \caption{Minimum \ce{CO2} adsorption energies in MOFs functionalized with the specified diamine as a function of diamine concentration. Diamine codes correspond to \cref{tab_diamines}.}
    \label{fig:diamine_scatter}
\end{figure*}

\clearpage
\FloatBarrier
\section{GCMC Placements}
\begin{figure}[htbp!]
    \centering
    \includegraphics[width=0.65\textwidth]{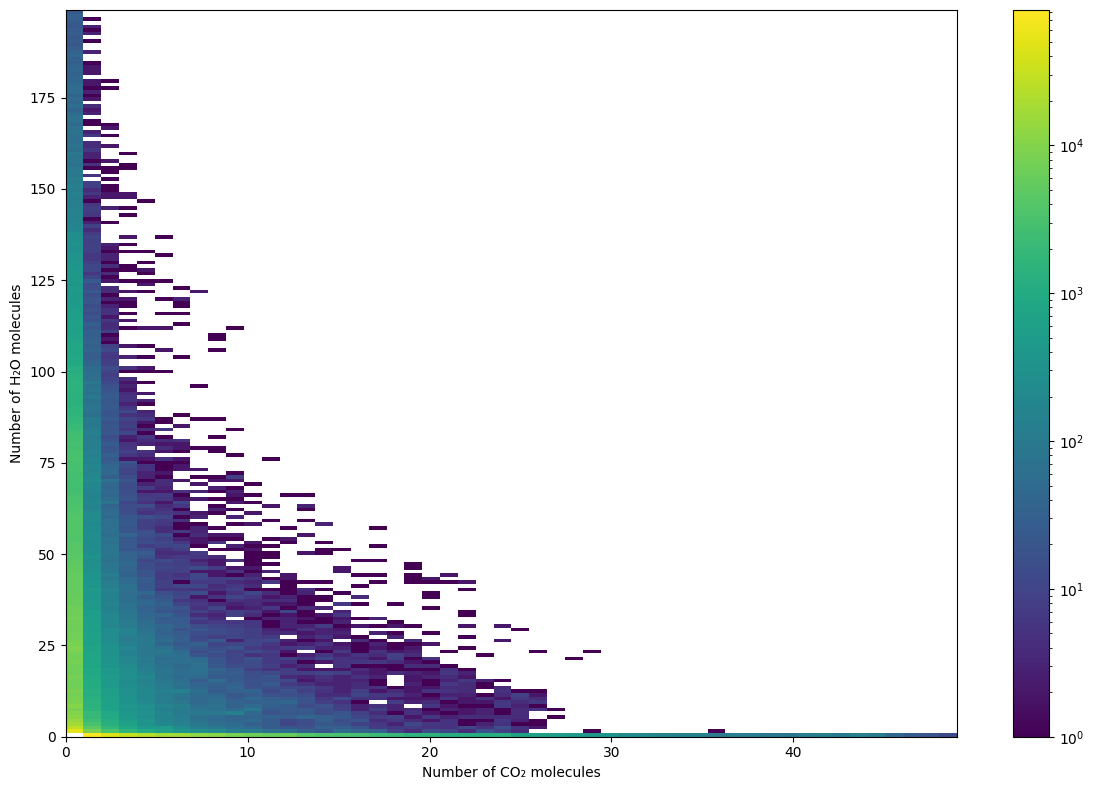}
    \caption{Two-dimensional histogram showing the distribution of \ce{CO2} and \ce{H2O} molecule counts in GCMC-generated configurations.}
    \label{fig:gcmc_hist}
\end{figure}

\clearpage
\FloatBarrier
\section{Synthetic MOF Properties}
\begin{figure}[htb!]
    \centering%
    \includegraphics[width=0.3\columnwidth]{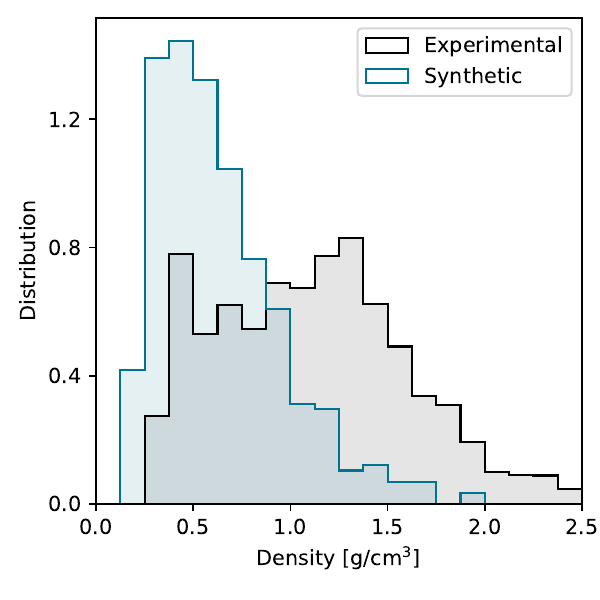}%
    \includegraphics[width=0.33\columnwidth]{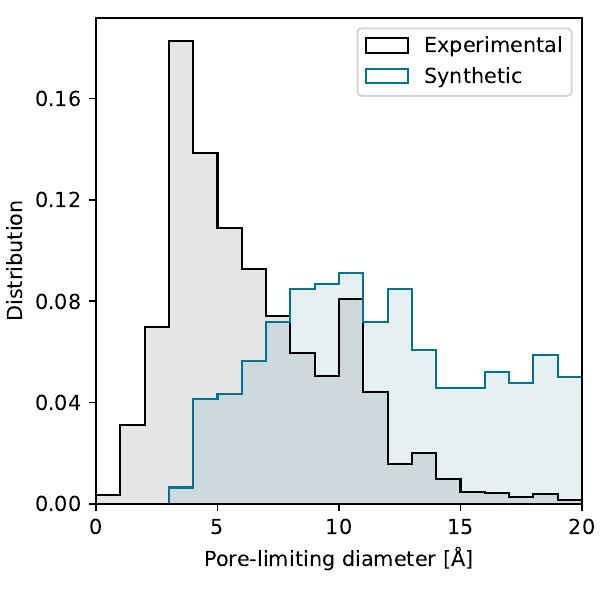}%
    \includegraphics[width=0.33\columnwidth]{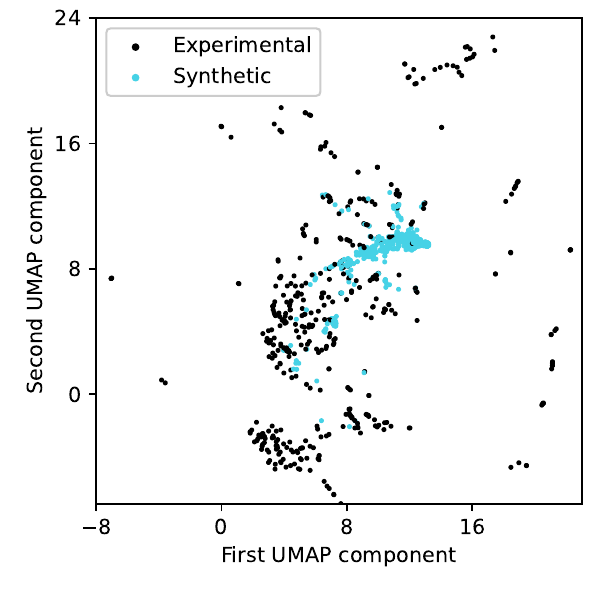}%
    \caption{Properties of synthetically generated MOFs (blue) compared to experimental structures (black). Left: density distribution. Middle: distribution of the PLD. Right: two-dimensional UMAP projection, where for legibility we only show 500 experimental MOFs uniformly sampled from the dataset.}%
    \label{fig:synthetic_mofs}%
\end{figure}

\begin{figure}[ht!]
    \centering
    \includegraphics[width=\textwidth]{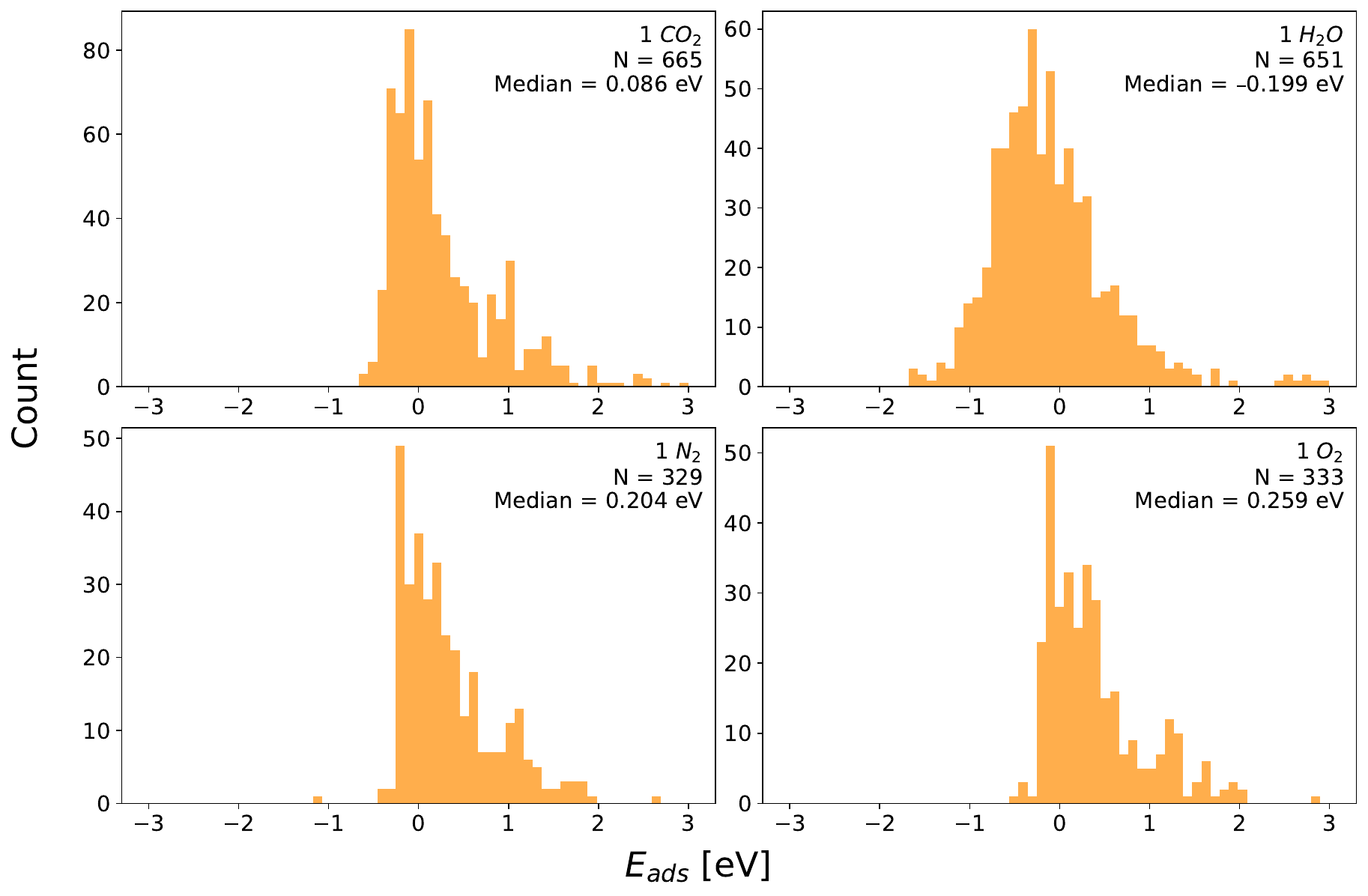}
    \caption{Distribution of single-molecule adsorption energies in synthetic MOFs split by adsorbate type.}
    \label{fig:synthetic_dist}
\end{figure}

\clearpage
\FloatBarrier
\section{MLIP Results}
\begin{table*}[!h]
    \caption{Hyperparameters and training details for the eSEN~\cite{fu2025learningsmoothexpressiveinteratomic} model trained on the ODAC25 dataset. Comma-separated values indicate pre-training, post-training parameters. The eSEN models were trained in two stages: first, a direct model with a maximum of 30 neighbors, and, subsequently, an energy-conserving model with up to 300 neighbors was trained. We limit the fine-tuning stage to structures with less than 350 atoms to reduce the GPU memory usage. Detailed descriptions of the eSEN model and architecture details are described in Fu et al.\cite{fu2025learningsmoothexpressiveinteratomic}.\label{tab:hyperparamaters}}
      \centering
\scalebox{0.8}{
\begin{tabular}{lc}
\toprule
Hyperparameters & eSEN-ODAC25\\ \midrule
Number of parameters & 50M\\
Maximum number of neighbors & 30, 300\\
Cutoff radius (\AA) & 6\\
Number of layers & 10\\
Number of sphere channels & 128\\
Number of edge channels & 128\\
Maximum degree $L_{max}$  & 4 \\
Maximum order $M_{max}$ & 2 \\
Distance function & gaussian \\
Number of distance basis & 128\\
Number of hidden channels & 128\\
Normalization type & rms\_norm\_sh \\
Activation type & gate \\
ff\_type & spectral \\
\multicolumn{1}{l}{} & \multicolumn{1}{l}{}\\
Number of GPUs & 64\\
Optimizer & AdamW\\
Learning rate scheduling & Cosine\\
Warmup epochs & 0.01\\
Warmup factor & 0.2\\
Maximum learning rate & $8 \times 10 ^{-4}$\\
Minimum learning rate factor & 0.01 \\
Gradient clipping norm threshold & 100\\
Weight decay & $1 \times 10 ^{-3}$\\
Batch size & 2000 atoms\\
Number of epochs & 4, 2 \\
Energy loss coefficient & 1, 20\\
Force loss coefficient & 20, 16\\
\bottomrule
\end{tabular}}
\end{table*}
\setlength{\tabcolsep}{4pt}
\begin{table*}[htb!]
\centering
\small
\caption{Performance breakdown of various models on the S2EF task by individual adsorbate configurations. Adsorption Energy MAE (EMAE) is reported in eV and Force MAE (FMAE) is reported in \eva. The UMA model was trained on a subset of ODAC25, while the two eSEN models were trained on the full and filtered versions of ODAC25. The best result in each column is shown in bold font.}
\label{tab:mlip_results_by_adsorbate}
\resizebox{\linewidth}{!}{
\begin{tabular}{lcccccccccc}
\toprule
\textbf{Model}
& \multicolumn{2}{c}{\textbf{\carbondioxide}}
& \multicolumn{2}{c}{\textbf{\water}}
& \multicolumn{2}{c}{\textbf{\carbondioxide+\water}}
& \multicolumn{2}{c}{\textbf{\nitrogen}}
& \multicolumn{2}{c}{\textbf{\oxygen}} \\
\cmidrule(lr){2-3}
\cmidrule(lr){4-5}
\cmidrule(lr){6-7}
\cmidrule(lr){8-9}
\cmidrule(lr){10-11}
& \emae & \fmae & \emae & \fmae & \emae & \fmae & \emae & \fmae & \emae & \fmae \\
\midrule
EquiformerV2 large\cite{ODAC23} & 0.117 & 0.025 & 0.232 & 0.025 & 0.129 & \textbf{0.024} & 0.219 & 0.036 & 0.242 & 0.048 \\
MACE~\cite{batatia2024foundationmodelatomisticmaterials} & 0.219 & 0.382 & 0.168 & 0.392 & 0.367 & 0.355 & 0.550 & 0.420 & 0.209 & 0.456 \\
MACE-DAC~\cite{mace_dac} & 0.106 & 0.301 & 0.149 & 0.317 & 0.252 & 0.272 & 0.380 & 0.322 & 0.284 & 0.368 \\
\midrule
UMA-Small 1.1~\cite{Wood1985} & 0.074 & 0.041 & 0.082 & 0.042 & 0.103 & 0.037 & 0.386 & 0.044 & 0.157 & 0.060 \\
eSEN (Full) & \textbf{0.061} & \textbf{0.024} & 0.075 & \textbf{0.022} & \textbf{0.081} & \textbf{0.024} & \textbf{0.081} & \textbf{0.028} & \textbf{0.068} & \textbf{0.044} \\
eSEN (Filtered) & 0.063 & 0.033 & \textbf{0.070} & 0.028 & 0.085 & 0.034 & 0.093 & 0.037 & 0.080 & 0.064 \\
\bottomrule
\end{tabular}
}
\end{table*}
\setlength{\tabcolsep}{1.4pt}

\clearpage
\FloatBarrier
\begin{figure*}[h]
    \centering
    \includegraphics[width=0.6\linewidth]{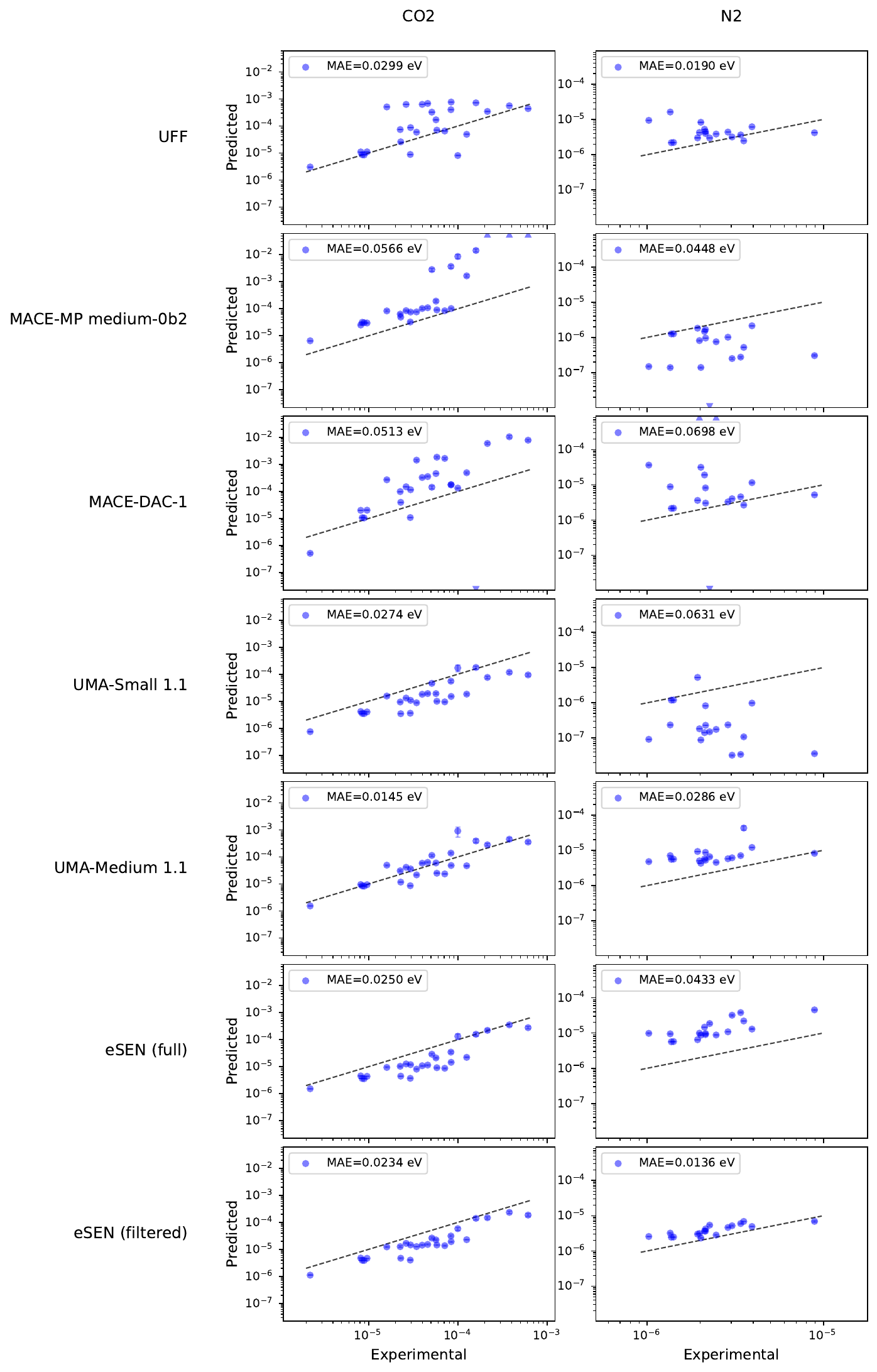}
    \caption{Log-log scatterplot of the Henry coefficient from experiment and from Widom insertion for the UFF baseline and different MLFFs, with the adsorbates \ce{CO2} and \ce{N2}.
    We show the mean absolute error between predictions and experimental values in eV, which is proportional to the logarithm of the Henry's constant~\citep{Yu2021} and the standard deviation of the predicted coefficient..
    }
    \label{fig:suppl-widom}
\end{figure*}

\clearpage
\FloatBarrier
\begin{table}[t!]
\centering
\footnotesize
\caption{Experimental Henry coefficients reported in mol/kg/Pa and eV.}
\label{tab:henry_exp}
\begin{tabular}{l l c c c}
\toprule
MOF & Adsorbate & Temp (K) & Henry coeff. (mol/kg/Pa) & Henry coeff. (eV) \\
\midrule
CALF20 & \ce{CO2} & 293 & 3.76e-04 & 0.199 \\
CALF20 & \ce{CO2} & 298 & 6.10e-04 & 0.190 \\
CALF20 & \ce{CO2} & 303 & 2.14e-04 & 0.221 \\
CALF20 & \ce{N2}  & 293 & 8.90e-06 & 0.294 \\
CALF20 & \ce{N2}  & 298 & 3.39e-06 & 0.323 \\
CALF20 & \ce{N2}  & 303 & 3.03e-06 & 0.332 \\
\midrule
CAU10  & \ce{CO2} & 296 & 4.59e-05 & 0.255 \\
CAU10  & \ce{CO2} & 298 & 3.98e-05 & 0.260 \\
CAU10  & \ce{CO2} & 303 & 1.60e-05 & 0.288 \\
CAU10  & \ce{N2}  & 298 & 1.35e-06 & 0.347 \\
\midrule
CaSQA  & \ce{CO2} & 298 & 1.59e-04 & 0.225 \\
CaSQA  & \ce{N2}  & 298 & 2.26e-06 & 0.334 \\
\midrule
FIQCEN & \ce{CO2} & 293 & 5.81e-05 & 0.246 \\
FIQCEN & \ce{CO2} & 295 & 7.12e-05 & 0.243 \\
FIQCEN & \ce{CO2} & 298 & 3.44e-05 & 0.264 \\
FIQCEN & \ce{N2}  & 293 & 2.87e-06 & 0.322 \\
FIQCEN & \ce{N2}  & 298 & 2.14e-06 & 0.335 \\
\midrule
KISXIU & \ce{CO2} & 298 & 9.62e-06 & 0.297 \\
KISXIU & \ce{N2}  & 298 & 1.93e-06 & 0.338 \\
\midrule
MIL160 & \ce{CO2} & 298 & 8.35e-05 & 0.241 \\
MIL160 & \ce{CO2} & 303 & 5.11e-05 & 0.258 \\
MIL160 & \ce{N2}  & 298 & 1.02e-06 & 0.354 \\
MIL160 & \ce{N2}  & 303 & 2.02e-06 & 0.342 \\
\midrule
MIL96  & \ce{CO2} & 298 & 9.96e-05 & 0.237 \\
\midrule
ORIWET & \ce{CO2} & 298 & 1.25e-04 & 0.231 \\
ORIWET & \ce{N2}  & 298 & 3.53e-06 & 0.322 \\
\midrule
PITYUN & \ce{CO2} & 298 & 2.22e-06 & 0.334 \\
\midrule
RUBTAK & \ce{CO2} & 298 & 2.95e-05 & 0.268 \\
RUBTAK & \ce{CO2} & 303 & 2.26e-05 & 0.279 \\
RUBTAK & \ce{N2}  & 298 & 1.98e-06 & 0.337 \\
RUBTAK & \ce{N2}  & 303 & 2.46e-06 & 0.337 \\
\midrule
RUBTAK-\ce{NH2} & \ce{CO2} & 298 & 5.68e-05 & 0.251 \\
RUBTAK-\ce{NH2} & \ce{N2}  & 298 & 2.11e-06 & 0.336 \\
\midrule
SAHYIK & \ce{CO2} & 296 & 2.93e-05 & 0.266 \\
SAHYIK & \ce{CO2} & 297 & 8.49e-06 & 0.299 \\
SAHYIK & \ce{CO2} & 298 & 8.92e-06 & 0.299 \\
SAHYIK & \ce{N2}  & 297 & 1.41e-06 & 0.345 \\
SAHYIK & \ce{N2}  & 298 & 1.37e-06 & 0.347 \\
\midrule
SUKXUS01 & \ce{CO2} & 298 & 8.17e-06 & 0.301 \\
\midrule
WIZMAV & \ce{CO2} & 298 & 2.29e-05 & 0.274 \\
\midrule
XITYOP & \ce{N2}  & 292 & 3.93e-06 & 0.313 \\
XITYOP & \ce{N2}  & 308 & 2.14e-06 & 0.346 \\
\midrule
XUPSAE & \ce{CO2} & 293 & 8.40e-05 & 0.237 \\
XUPSAE & \ce{CO2} & 298 & 2.63e-05 & 0.271 \\
\bottomrule
\end{tabular}
\end{table}

\clearpage
\FloatBarrier
\section{MOF Deformation}

\setlength{\tabcolsep}{4pt}
\begin{table*}[htb!]
\centering
\small
\caption{Mean absolute errors in eV for energies relative to DFT calculations from ref. \cite{Brabson2025}. CHGNet and EqV2 results are adapted from that work for comparison.}
\label{tab:deform_results}
\begin{tabular}{lccc}
\toprule
\textbf{Model} & \textbf{Adsorption energy} & \textbf{Interaction energy} & \textbf{MOF deformation energy} \\
\midrule
eSEN-ODAC25-Full & 0.125 & 0.118 & 0.096 \\
eSEN-ODAC25-Filtered & \textbf{0.087} & \textbf{0.103} & \textbf{0.054} \\
CHGNet \cite{deng2023chgnet-39f} & 0.124 & 0.163 & 0.065 \\
EqV2-ODAC23 \cite{ODAC23} & 0.191 & -- & -- \\
\bottomrule
\end{tabular}
\end{table*}
\setlength{\tabcolsep}{1.4pt}

\begin{figure*}[h]
    \centering
    \includegraphics[width=\linewidth]{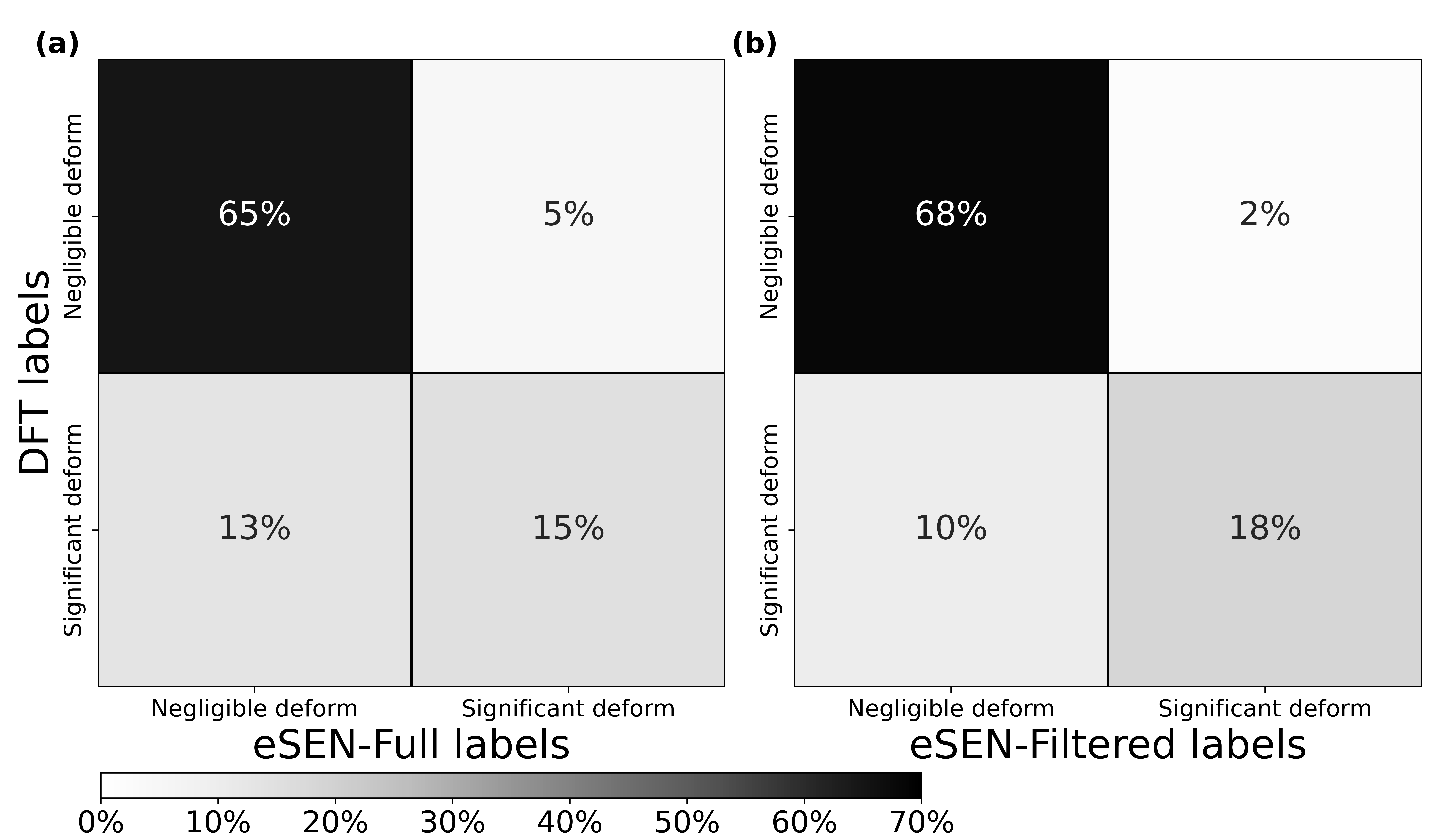}
    \caption{Confusion matrices for determining MOF deformation class using (a) eSEN-Full and (b) eSEN-Filtered models for the 59 MOF+adsorbate systems from ref. \cite{Brabson2025}. True positives and true negatives are located in the upper left and lower right of each matrix, respectively.}
    \label{fig:confusion}
\end{figure*}

\end{document}